\documentclass[10pt]{article}
\usepackage{fullpage} 
\usepackage[section]{placeins}
\usepackage[superscript]{cite}
\usepackage[colorlinks = true,
linkcolor = blue,
urlcolor  = blue,
citecolor = blue,
anchorcolor = blue]{hyperref}

\usepackage{authblk}
\usepackage{graphicx}
\usepackage{amsmath}
\usepackage[dvipsnames]{xcolor}
\usepackage{bbm}

\makeatletter
\renewcommand\@biblabel[1]{\textbf{#1.}\hfill}
\makeatother

\begin{document}
\title{Synchronization, Collective Oscillations, and Information Flow in Duplex Networks}
\author[1]{Ali Seif}
\author[1]{Mina Zarei\thanks{Corresponding author. Email: \href{mailto:mina.zarei@iasbs.ac.ir}{mina.zarei@iasbs.ac.ir}}}
\affil[1]{Institute of Advanced Studies in Basic Sciences (IASBS), Department of Physics, Zanjan, 45137-66731, Iran.}
\vspace{-1em}
\date{\today}
\maketitle 
\pagenumbering{arabic} 


\begin{abstract}
	\noindent
	{In many real-world systems, partial synchronization is the dominant dynamical regime and, in systems such as the brain, is often accompanied by collective oscillations in which multiple overlapping modes interact to produce complex rhythmic activity. Here, we investigate duplex networks with reactive interlayer links, where full synchronization cannot be achieved. We show that when interlayer frequency differences between mirror nodes are uniformly distributed with sufficient width, the network self-organizes into collective macroscopic oscillations composed of multiple interacting modes. By linking macroscopic phase transitions to microscopic directed information transfer between nodes, we uncover the mechanisms underlying the emergence of these multimodal dynamics.}
	\vspace{1em} 
	
	\noindent keywords: {Multiplex networks, Explosive synchronization, Interlayer frustration, Interlayer frequency mismatches, Collective oscillations, Transfer entropy}
\end{abstract}


\section{Introduction}
\addcontentsline{toc}{section}{Introduction}
\label{Introduction}

Synchronization is a fundamental phenomenon in many natural and engineered systems, ranging from neuronal populations in the brain to power grids and social networks~\cite{strogatz2004sync,boccaletti2006complex,arenas2008synchronization}. The transition from incoherence to collective synchronization has been extensively studied, showing how coherent behavior can emerge either gradually (continuous transition) or abruptly (discontinuous transition) as the coupling strength increases~\cite{kumar2021explosive,strogatz2000kuramoto,gomez2011explosive}. Importantly, synchronization is not merely a macroscopic outcome but arises from ongoing dynamical interactions among oscillators. These interactions produce directed exchanges of influence that shape the collective state of the system. By quantifying this directed information flow through transfer entropy~\cite{schreiber2000measuring,vicente2011transfer,lizier2014jidt}, enables us to link microscopic interactions between pairs of nodes to the global synchronization pattern. In this sense, the emergence and nature of collective order are closely related to how information is structured and redistributed within the network.

Previous studies have further shown that total information transfer is typically low in both fully incoherent and fully synchronized states, reaching a maximum near the synchronization transition~\cite{vicente2011transfer}. This highlights the strong connection between dynamical criticality and enhanced information exchange in complex systems. Moreover, the direction of information flow is not random: it often originates from nodes with large absolute natural frequencies or from highly connected hubs and propagates toward more peripheral nodes~\cite{sancristobal2014role,pariz2018high,novelli2020deriving}. However, this hierarchy is not universal. Changes in system parameters can reorganize the direction of information flow; for example, time delay has been shown to reverse or redistribute causal influences among nodes~\cite{pariz2021transmission}. Therefore, mechanisms such as time delay, frustration, and interlayer interactions can significantly reshape information flow patterns and, consequently, influence the macroscopic dynamics and the nature of synchronization phase transitions.

Prior research has revealed that introducing reactive interlayer interactions in a duplex network—by setting the interlayer frustration to $\frac{\pi}{2}$—can fundamentally alter the nature of synchronization within each layer~\cite{jalan2019explosive,jalan2019inhibition,seif2025double}. In particular, when a nonzero average frequency difference exists between mirror nodes across the layers, the intralayer synchronization transition can change from continuous to discontinuous (explosive)~\cite{jalan2019explosive}. This indicates that the combination of reactive coupling and interlayer frequency mismatch plays a crucial role in reshaping the collective dynamics and determining the nature of the synchronization transition within each layer.

In our previous work, we demonstrated that in duplex networks with reactive interlayer links, collective dynamics are determined not only by the average interlayer frequency mismatch but also by the arrangement of frequencies across mirror nodes~\cite{seif2025double}. Specifically, we showed that two configurations with the same average interlayer frequency difference can lead to qualitatively different behaviors. In one case, the system undergoes a single explosive synchronization transition with a clear hysteresis loop. In another case, the system displays a double explosive transition accompanied by oscillations in the order parameter, where different dynamical modes coexist and overlap. These results highlight that the detailed distribution and arrangement of interlayer frequency mismatches play a crucial role in shaping the emergent collective behavior.

Collective oscillations, often comprising multiple superimposed frequency modes, are central to many systems, from neural networks and biochemical circuits to power grids, climate systems, and social or ecological networks~\cite{kuramoto2002coexistence,ott2008low,abrams2008solvable,motter2013spontaneous,dijkstra2013nonlinear,elowitz2000synthetic}. Transitions between oscillatory regimes—or the loss of rhythmicity—often signal critical changes, such as functional breakdown or phase transitions. In the brain, for example, neural populations exhibit diverse rhythms across frequency bands (alpha, beta, gamma) that coordinate information exchange and cognitive processes~\cite{buzsaki2004neuronal,buzsaki2006rhythms,fries2005mechanism,miller2018working,canolty2010functional}. Multiple oscillatory modes can coexist and interact, producing complex macroscopic temporal patterns. Understanding how theoretical models generate such collective oscillations is therefore crucial for uncovering the mechanisms of emergent dynamics and for predicting, controlling, and optimizing complex systems across disciplines.

Despite their ubiquity, the mechanisms underlying collective oscillations remain an active area of research. The Kuramoto model offers a robust framework for studying periodic synchronization~\cite{kuramoto1984chemical,kuramoto2005self,acebron2005kuramoto}. In partially synchronized regimes, networks can exhibit sustained oscillatory behavior; for instance, fully connected networks with bimodal frequency distributions can support limit-cycle solutions~\cite{martens2009exact}. Additionally, coupled second-order Kuramoto oscillators may display periodic dynamics depending on parameters such as inertia and time delays~\cite{mahdavi2025synchronization}.

On the other hand, many real-world systems are naturally multiplex networks, where nodes interact through multiple types of connections simultaneously~\cite{boccaletti2014structure,kivela2014multilayer,bianconi2018multilayer}. Each layer can represent a different type of interaction, with its own structure and dynamics. Examples include neurons connected by electrical and chemical synapses or social ties across professional and online networks. These multiplex structures cannot be fully captured by single-layer networks, as interlayer interactions strongly shape collective behavior and synchronization~\cite{de2013mathematical,gomez2013diffusion}.

In this study, building on our previous work, we investigate the dynamics of duplex networks by extending the Kuramoto model to multiplex systems. Specifically, we examine how interlayer interactions—ranging from dissipative to reactive (frustrated) links—and the distribution of interlayer frequency mismatches reshape collective dynamics. We focus on their effects on phase transitions, the emergence of oscillatory behavior, and the patterns of information transfer both within and between layers.

The paper is organized as follows. In Section~\ref{Methods}, we introduce the models and methods, describing the duplex network structure, the extended Kuramoto dynamics, and the transfer entropy framework used to quantify directed information flow. Section~\ref{Results} presents the results, including the synchronization dynamics under different interlayer interaction types and frequency-mismatch distributions, as well as the corresponding analysis of information transfer within and between layers. Finally, Section~\ref{Discussion} provides a discussion of the implications of our findings, highlighting how interlayer interactions and frequency arrangements influence collective oscillations, phase transitions, and information flow in multiplex networks.


\section{Methods}

\label{Methods}
This section describes the methodological framework employed in this study.


\subsection{Network Model and Dynamical Properties}
\label{NetworkModel}
The system consists of coupled oscillators whose behavior is shaped by network structure, dynamical rules, and the assignment of natural frequencies. Together, these elements define the structural and dynamical landscape of the network, providing a framework for analyzing phase transitions and information transfer.


\vspace{0.5em}
\subsubsection{Network Structure}
\label{NetworkStructure}

\begin{figure}[!ht]
		\centering
		\includegraphics[width=0.46\linewidth]{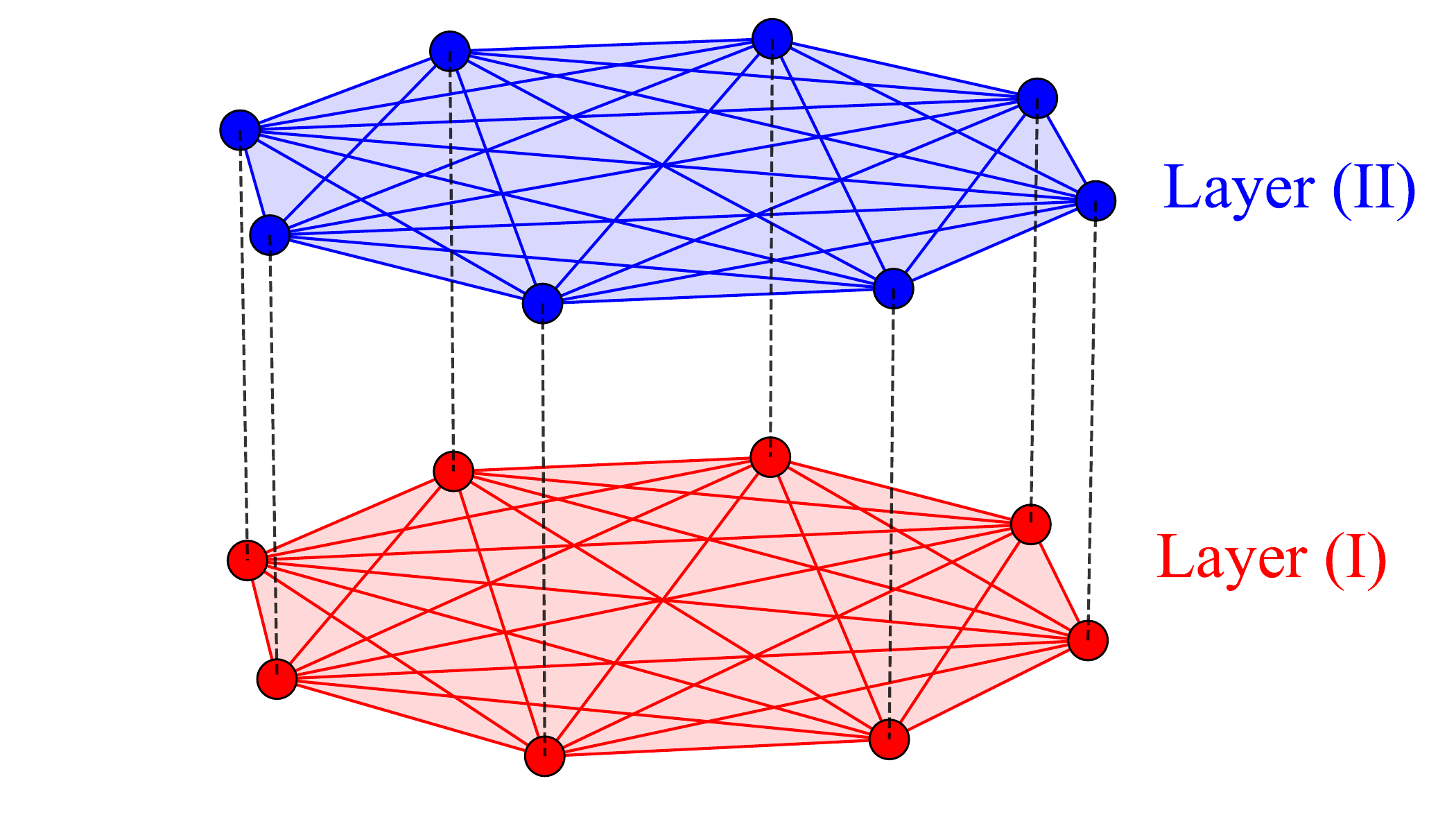}
		\caption{A schematic illustration of a duplex network with two fully connected layers. Solid lines show the intralayer connections, while dashed lines denote the interlayer couplings that connect each node to its mirror node in the other layer.}
		\label{fig:fig1}
	\end{figure}

The network considered here is a two-layer multiplex structure in which each layer forms a fully connected graph. Nodes within a layer represent individual oscillators, while interlayer couplings link each node to its mirror counterpart in the other layer. This configuration enables both intra- and interlayer interactions, providing a controlled framework for investigating how network structure and multiplexing influence dynamical behavior. A schematic of the duplex network is shown in Fig.~\ref{fig:fig1}.

\vspace{0.5em}
\subsubsection{Network Dynamics}
\label{NetworkDynamics}

Having defined the duplex network structure, we now turn to the dynamical mechanisms governing the temporal evolution of the oscillators.Each node is modeled as a phase oscillator whose dynamics are described by an extended Kuramoto equation. This framework allows us to systematically investigate how intra- and interlayer couplings jointly shape the emergence of global synchronization and phase coherence across the network.

The classical Kuramoto model was originally introduced to explain spontaneous synchronization in ensembles of globally coupled phase oscillators~\cite{kuramoto1984chemical,acebron2005kuramoto}. Over time, this framework has been extended to incorporate complex and real network features, enabling the study of how factors such as network topology~\cite{strogatz2000kuramoto}, multilayer interactions~\cite{ghosh2025transitions}, time delays~\cite{yeung1999time,ziaeemehr2020frequency}, phase frustration~\cite{sakaguchi1986soluble}, and the distribution of natural frequencies influence the emergence of collective dynamical behavior~\cite{kuramoto1984chemical,seif2025double}.

As outlined above, we consider a duplex Kuramoto model consisting of two fully connected layers, each comprising $N$ oscillators. Oscillators within the same layer interact via all-to-all coupling with strength $\sigma$, while corresponding mirror nodes in different layers are connected through interlayer couplings of strength $\lambda$. To account for the role of multiplexing in shaping the system’s dynamics, we include a constant phase-lag parameter $\alpha$ in the interlayer coupling, which captures frustration effects that may significantly influence synchronization and dynamical stability.

Accordingly, the phase dynamics of oscillator $i$ in layer~$I$ and its corresponding mirror node in layer~$II$ are described by the following set of coupled differential equations~\cite{kumar2021explosive}:

\begin{subequations}
\begin{flalign}
&\dot{\theta}_i^{I} = \omega_i^I + \frac{\sigma}{N} \sum_{j=1}^N \sin \left(\theta_j^I - \theta_i^I\right) + \lambda \sin \left(\theta_i^{II} - \theta_i^I + \alpha\right), \label{eq:1-a} \\
&\dot{\theta}_i^{II} = \omega_i^{II} + \frac{\sigma}{N} \sum_{j=1}^N \sin \left(\theta_j^{II} - \theta_i^{II}\right) + \lambda \sin \left(\theta_i^I - \theta_i^{II} + \alpha\right), \label{eq:1-b}
\end{flalign}
\label{eq:1}
\end{subequations}
where $\theta_i^{I(II)}$ and $\omega_i^{I(II)}$ denote the instantaneous phases and natural frequencies, respectively. 
The terms on the right-hand side of Eq.~\ref{eq:1} represent, from left to right, the intrinsic oscillator dynamics, the intralayer coupling, and the interlayer interaction with phase lag~$\alpha$.
Collective synchronization in each layer is quantified by the standard Kuramoto order parameter:
\begin{equation}
    r^{I(II)} e^{i \psi^{I(II)}} = \frac{1}{N} \sum_{j=1}^{N} e^{i \theta_j^{I(II)}},
    \label{eq:2}
\end{equation}
where $r^{I(II)} \in [0,1]$ quantifies the degree of phase coherence in layer~I~(II), and $\psi^{I(II)}$ denotes the global (mean-field) phase associated with the collective dynamics of that layer. 
Here, $r^{I(II)} = 1$ corresponds to complete synchronization, whereas $r^{I(II)} \approx 0$ indicates an incoherent state.

In addition to the global order parameters that characterize synchronization at the layer and interlayer levels, it is also informative to examine the instantaneous pairwise phase relationships between oscillators.
To this end, we define a \textit{phase similarity} matrix:
\begin{equation}
	S_{i j}^{I(II)}(t)= \cos \left(\theta_i^{I(II)}(t)-\theta_j^{I(II)}(t)\right), 
	\label{eq:3}
\end{equation}
which quantifies the instantaneous local phase coherence between oscillators $i$ and $j$ within layer~I or~II at time~$t$.
Here, $S_{ij}^{I(II)} = 1$ corresponds to perfect phase alignment, while $S_{ij}^{I(II)} = -1$ indicates complete anti-phase behavior.
This measure captures instantaneous correlations between oscillators, providing insight into the detailed dynamics of the system, such as which groups of nodes synchronize with one another.


\vspace{0.5em}
\subsubsection{Natural Frequency Configurations in the Duplex Network}

\label{NaturalFrequency}
The distribution of natural frequencies plays a central role in shaping synchronization dynamics in networks of coupled oscillators. In duplex networks, prior studies have shown that synchronization is determined not only by the mean frequency mismatch between layers but also by the manner in which natural frequencies are distributed and assigned to nodes and their mirror counterparts. Even subtle differences in the organization of natural frequencies can thus lead to qualitatively distinct collective behaviors~\cite{seif2025double,kumar2021explosive}.

Here, we investigate how the distribution of frequency differences between mirror nodes influences network dynamics. To enable meaningful comparisons across different distributions, we keep the average frequency difference between layers constant, which requires defining a measure of the interlayer frequency mismatch.

To quantify the mismatch between mirror nodes, we define the \emph{mirror-node frequency difference} for each node $i$ as
\begin{equation}
	\delta \omega_i = \omega_i^{II} - \omega_i^{I}, \quad i = 1,\dots,N,
	\label{eq:4}
\end{equation}
where $\omega_i^{I}$ and $\omega_i^{II}$ denote the natural frequencies of node $i$ in layers~I and~II, respectively. The distribution of $\delta \omega_i$ across all mirror pairs provides a statistical characterization of interlayer heterogeneity, which directly influences duplex network dynamics.

To characterize the average frequency mismatch between layers in a normalized and comparable way, we use the following measure~\cite{kumar2021explosive}:
\begin{equation}
	\Delta \omega = \frac{\sum_{i=1}^{N} \left| \delta \omega_i \right|}{\sum_{i=1}^{N} \left| \omega_i^{I} \right| + \sum_{i=1}^{N} \left| \omega_i^{II} \right|}.
	\label{eq:5}
\end{equation}
This dimensionless measure satisfies $0 \leq \Delta \omega \leq 1$. The limit $\Delta \omega = 0$ corresponds to perfect inter-layer frequency matching, while $\Delta \omega = 1$ represents the maximal mismatch, in which mirror-node frequencies are equal in magnitude but opposite in sign. By keeping $\Delta \omega$ constant, we can compare different interlayer frequency-mismatch distributions.


\vspace{0.5em}
\subsection{Information-Theoretic Framework and Estimation Methods}
\label{Information-Theoretic}
After establishing the structural and dynamical rules governing the multiplex Kuramoto system, we seek a principled way to quantify how oscillators influence one another through the information they exchange as the network moves toward or away from synchronization. Information theory provides a natural framework for this purpose~\cite{shannon1948mathematical}. Unlike traditional dynamical measures that focus on phase coherence or frequency locking, information-theoretic quantities reveal how the state of one oscillator reduces uncertainty about another, capturing subtle directional interactions that may be hidden from the order parameter alone.


\vspace{0.5em}
\subsubsection{Transfer Entropy as a Measure of Directed Information Flow}
\label{Transfer Entropy}
Shannon entropy lies at the heart of this framework, providing a quantitative measure of the uncertainty inherent in a random variable. For a random variable $X$ with probability distribution $p(x)$, the entropy is defined as
\begin{equation}
	H(X) = - \sum_x p(x) \log p(x),
	\label{eq:6}
\end{equation}
The sum is taken over all possible values that the random variable $X$ can assume, measuring the average amount of information required to describe the outcome of $X$. The entropy satisfies $0 \leq H(X) \leq \log_2 n$, where $n$ denotes the number of possible states of $X$. 

For two variables, $X$ and $Y$, their statistical dependence is captured by the \emph{mutual information} (MI)~\cite{cover1999elements}:

\begin{equation}
I(X;Y) = \sum_{x,y} p(x,y) \log \frac{p(x,y)}{p(x)p(y)},
\label{eq:7}
\end{equation}
"where $p(x,y)$ is the joint probability distribution of $X$ and $Y$, and $p(x)$ and $p(y)$ are the corresponding marginal distributions of $X$ and $Y$. This quantifies how knowledge of $Y$ reduces the uncertainty of $X$. Equivalently, MI can be expressed in terms of entropies as
\begin{equation}
I(X;Y) = H(X) - H(X|Y),
\label{eq:8}
\end{equation}
where $H(X|Y)$ denotes the conditional entropy of $X$ given $Y$. While mutual information $I(X;Y)$ effectively measures statistical dependence, it is symmetric and therefore cannot capture the directionality of influence between variables. To determine which oscillators “drive” others within the network, it is necessary to employ a metric that is sensitive to temporal influence. 

Synchronization is inherently a dynamical process, and understanding it requires quantifying how the past state of one oscillator shapes the future state of another. \emph{Transfer entropy} (TE) captures precisely this type of directional, time-asymmetric influence~\cite{schreiber2000measuring}.

In our context, TE tells us whether knowing the past phase trajectory of oscillator $Y$ improves the prediction of the future phase of oscillator 
$X$, above and beyond what $X$’s own past already provides. This is particularly relevant near the synchronization threshold~\cite{lizier2008local,barnett2012transfer,lizier2014framework}, where oscillators interact irregularly and causal influences reorganize.

Formally, the transfer entropy from $Y$ to $X$ is defined as
\begin{equation}
TE_{Y \rightarrow X} = \sum_{x_{t+1},\,x_t,\,y_t}
p(x_{t+1}, x_t, y_t)
\log \frac{p(x_{t+1} \mid x_t, y_t)}{p(x_{t+1} \mid x_t)},
\label{eq:9}
\end{equation}

Here, $x_{t+1}$ represents the future state of variable $X$, while $x_t$ and $y_t$ denote the current states of $X$ and $Y$, respectively. The term $p(x_{t+1}, x_t, y_t)$ is the joint probability distribution of these three states, and $p(x_{t+1} \mid x_t, y_t)$ and $p(x_{t+1} \mid x_t)$ are the corresponding conditional probabilities. Transfer entropy $TE_{Y \rightarrow X}$ thus quantifies the reduction in uncertainty about the future state of $X$ due to knowledge of the current state of $Y$, beyond what is already contained in the current state of $X$ itself. 

TE is indeed equivalent to the conditional mutual information~\cite{barnett2009granger,vicente2011transfer}, $TE_{Y \rightarrow X} = I(X_{t+1}; Y_t \mid X_t)$, 
and takes a nonzero value only if the past of $Y$ helps predict the future of 
$X$ beyond what is predictable from $X$’s own history. Transfer entropy (TE) thus serves as a probe of causal interactions between oscillators, enabling us to map the flow of predictive, directional information both within and across layers of the duplex network~\cite{battiston2014structural}.


\vspace{0.5em}
\subsubsection{Estimating Transfer Entropy Using JIDT}
\label{Estimation}

All transfer entropy analyses were performed using the Java Information Dynamics Toolkit (JIDT)~\cite{lizier2014jidt}, an open-source package designed for estimating information-theoretic measures in complex dynamical systems. Accurate computation of transfer entropy requires reliable estimation of the joint and conditional probability distributions, $p(x_{t+1}, x_t, y_t)$, which can be challenging for finite datasets, particularly in high-dimensional or continuous-valued systems, due to limited sampling and the curse of dimensionality. To address this, JIDT implements a multivariate extension of the Kraskov–Stögbauer–Grassberger (KSG) estimator~\cite{kraskov2004estimating,vicente2011transfer}, a nonparametric and data-efficient method for estimating mutual information and related quantities, including transfer entropy.

In our study, the time series correspond to the phases of coupled oscillators, which are inherently periodic~\cite{lobier2014phase}, necessitating an estimator capable of handling circular distributions and nonlinear dependencies. The multivariate KSG estimator is well suited for this purpose, as it nonparametrically captures the local structure of the joint probability space, making it effective for analyzing nonlinear dynamical systems such as the Kuramoto model.

To evaluate the statistical significance of the estimated information transfer, we generated \textit{time-shift surrogate data}~\cite{theiler1992testing,schreiber2000surrogate}, in which the source time series is randomly shifted in time. This disrupts its temporal alignment with the target while preserving amplitude distribution and autocorrelation, providing a null model representing the absence of directed interactions. Transfer entropy was computed for each surrogate realization, $\widehat{TE}_{Y \to X}^{(s)}$, and the significance of the observed value, $\widehat{TE}_{Y \to X}^{\text{obs}}$, was assessed using a one-sided test:

\begin{equation}
	p = \frac{1}{S} \sum_{s=1}^{S} \mathbbm{1}\!\left( \widehat{TE}_{Y \to X}^{(s)} \ge \widehat{TE}_{Y \to X}^{\text{obs}} \right),
	\label{eq:10}
\end{equation}

where $\mathbbm{1}(\cdot)$ denotes the indicator function. The resulting p-value quantifies the likelihood that the observed transfer entropy could arise under the null hypothesis of no directed interaction. We applied a significance threshold of $p < 0.05$, allowing rigorous identification of statistically meaningful information transfer between oscillators while mitigating finite-sample biases.

	
\vspace{0.5em}
\section{Results}

\label{Results}

The primary objective of this study is to investigate how frequency mismatches between mirror nodes in a duplex network influence phase transitions, the emergence of oscillatory behavior at the population level, and the mechanisms governing information transfer within and between layers.

To systematically explore these effects, the results are organized into two main parts. First, we investigate the synchronization transitions of the duplex network and their modulation by frequency mismatch. We then focus on the information-theoretic analysis, examining how directed information flow varies with intralayer coupling strength within and between layers, particularly in the phase transition regime.


\subsection{Interlayer Frequency Mismatches and Oscillatory Synchronization}

\label{ResSynchronization}

	\begin{figure}[!ht]
	\centering
	\includegraphics[width=1.0\linewidth]{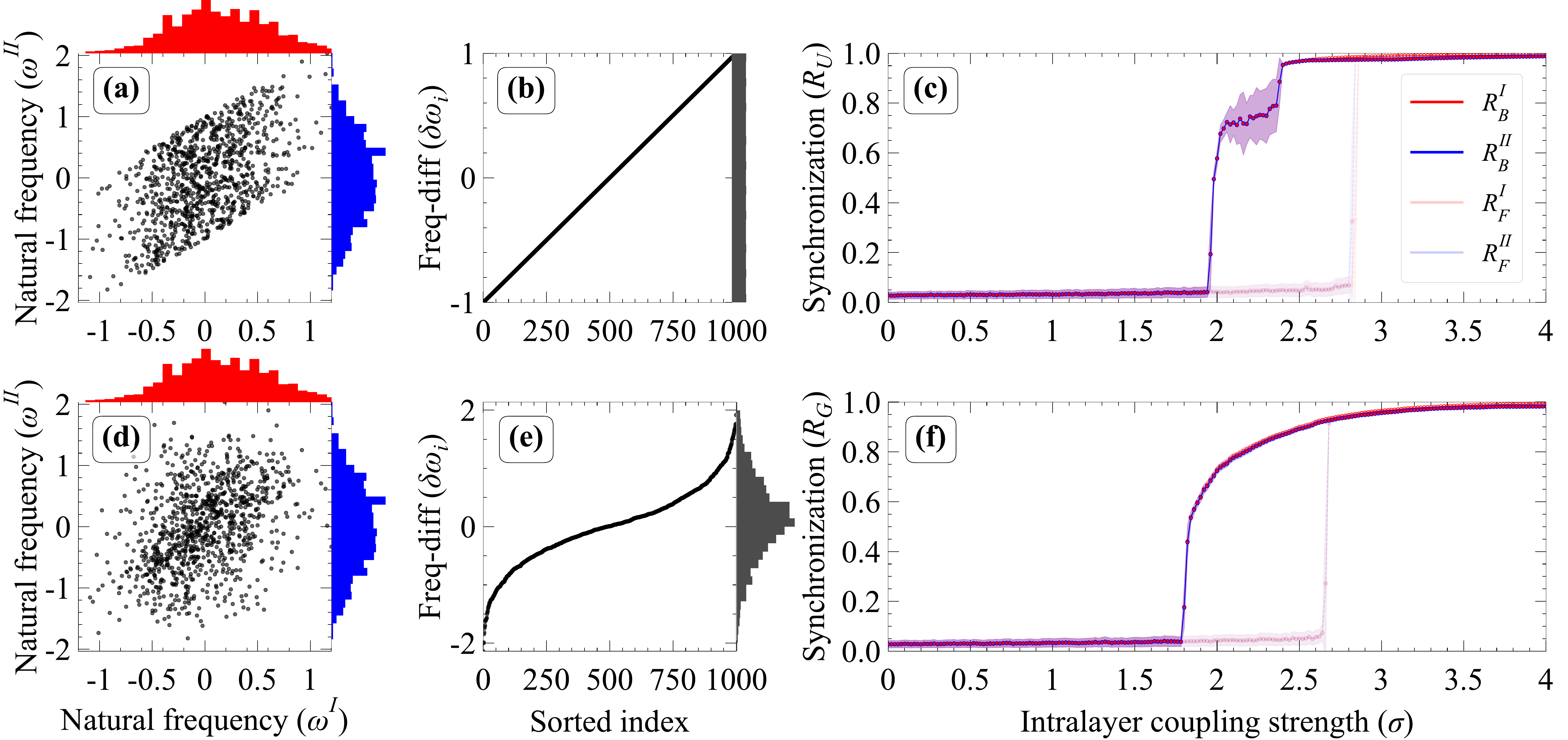}
	\caption{Phase transition behavior in two globally similar frequency configurations with distinct interlayer frequency-mismatch distributions. Panels (\textbf{a--c}) correspond to a uniform (step-function) distribution of mirror-node frequency differences, while panels (\textbf{d--f}) correspond to a Gaussian distribution. In both cases, layers~I and~II (each with $N = 1000$ nodes) share identical natural frequencies across the two configurations, although the frequencies in each layer are drawn from different Gaussian distributions. The interlayer interaction includes a frustration parameter $\alpha = \frac{\pi}{2}$, and both configurations have the same average frequency mismatch, $\Delta\omega = 0.56$; thus, any differences in behavior arise solely from the specific pairing of mirror nodes. 	
		The \textbf{first column} shows the natural frequencies of layer~II nodes versus those of layer~I, with marginal plots indicating their respective distributions (red for layer~I and blue for layer~II). The \textbf{second column} displays the mirror-node frequency differences (Freq-diff, $\delta \omega_i$) as a function of node index, ordered from low to high values, with the corresponding distributions shown alongside each panel (uniform in the top row and Gaussian in the bottom row). The \textbf{third column} presents the phase transition along the forward (light lines) and backward (dark lines) paths, with red and blue curves representing layers~I and~II, respectively; shaded regions denote the temporal standard deviation.
	}
	\label{fig:fig2}
\end{figure}
\begin{figure}[!ht]
	\centering
	\includegraphics[width=1\linewidth]{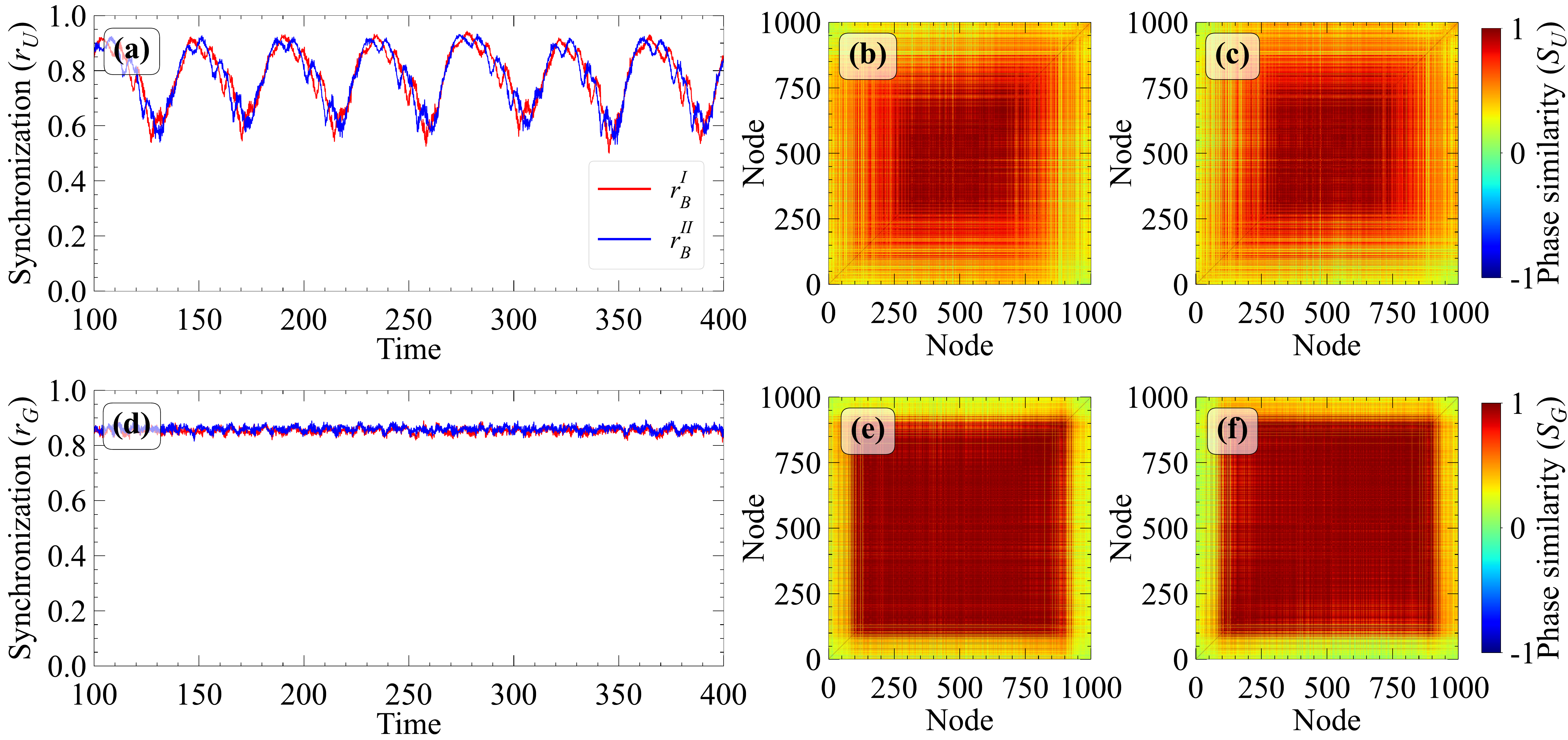}
	\caption{Synchronization dynamics for two frequency configurations with step-function and Gaussian mirror-node frequency differences, exhibiting oscillatory and non-oscillatory order-parameter behavior. The first row corresponds to the uniform (step-function) distribution, while the second row corresponds to the Gaussian distribution. Panels (\textbf{a}) and (\textbf{d}) show the time evolution of the synchronization order parameter at $\sigma = 2.34$, with blue and red curves representing layers~I and~II, respectively. Panels (\textbf{b}) and (\textbf{c}) show the corresponding time-averaged phase similarity matrices for layers~I and~II in the step-function case, while panels (\textbf{e}) and (\textbf{f}) show the mean phase similarity matrices for layers~I and~II under the Gaussian distribution. Nodes in the phase similarity matrices are ordered according to $\delta \omega_i$ from low to high values. The network structure and frequency arrangement are the same as in the previous figure.
	}
	\label{fig:fig3}
\end{figure}

\begin{figure}[!ht]
	\centering
	\includegraphics[width=0.75\linewidth]{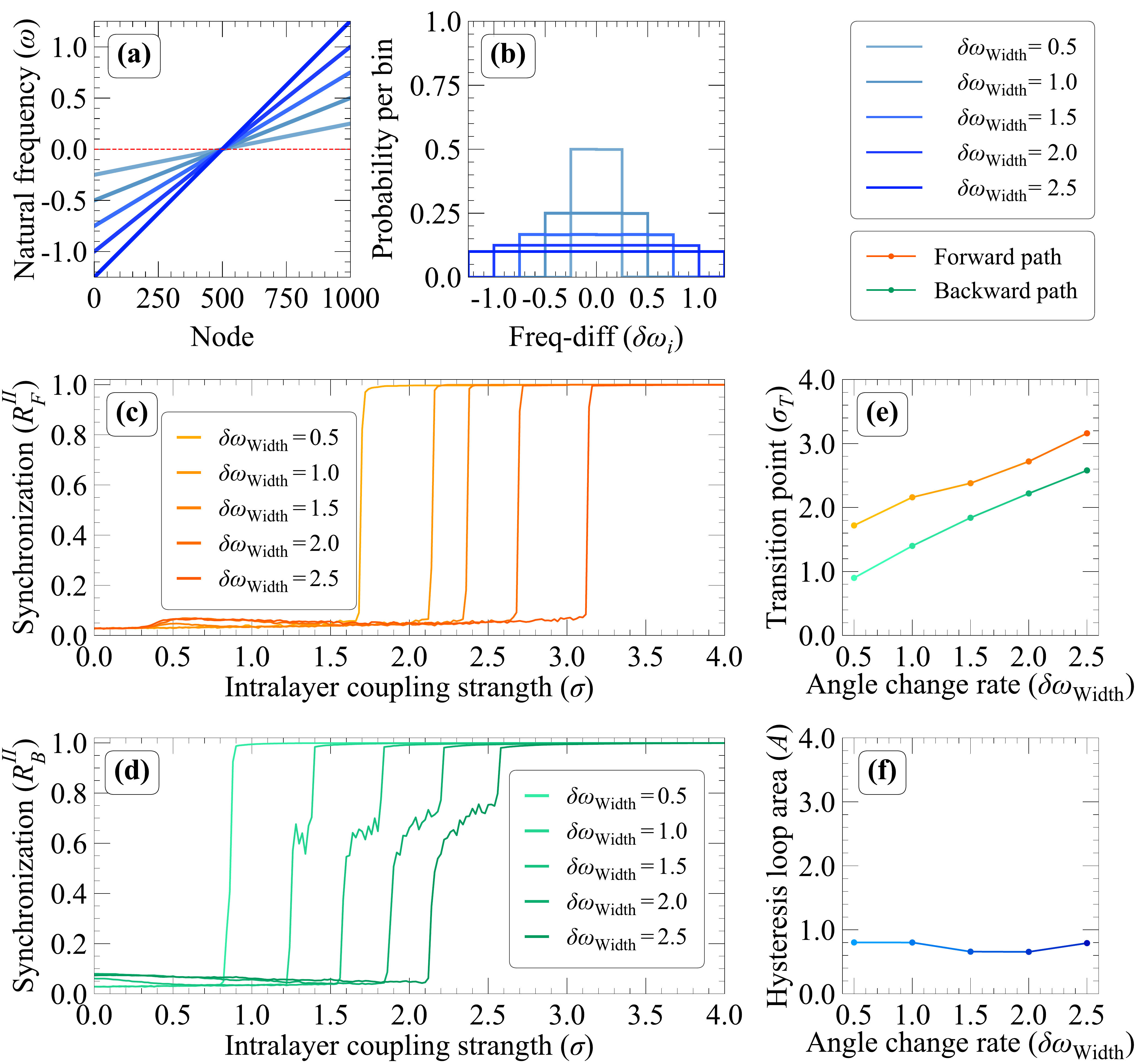}
	\caption{Synchronization dynamics for configurations with uniform (step-function) mirror-node frequency differences of systematically varying widths. The first row illustrates the method for arranging frequencies with a uniform mirror-node frequency-mismatch distribution. \textbf{(a)} Natural frequencies of nodes versus node index, with red and blue lines representing layers~I and II, respectively. \textbf{(b)} Probability density of mirror-node frequency differences across the layers, with the color gradient from light to dark blue indicating increasing widths of the distribution. Node indices are ordered according to $\delta \omega_i$ from low to high values. \textbf{(c, d)} Synchronization transition in layer~II of a duplex network of two fully connected layers ($N = 1000$ nodes, $\alpha = \frac{\pi}{2}$) as the intralayer coupling is varied, shown for forward (orange) and backward (green) paths. Light to dark colors indicate increasing widths of the interlayer frequency-mismatch distribution. \textbf{(e)} Transition couplings $\sigma_T$ for forward (orange) and backward (green) paths versus distribution width. \textbf{(f)} Hysteresis loop area $A$ between forward and backward paths versus distribution width.	}
	\label{fig:fig5}
\end{figure}

\begin{figure}[!ht]
	\centering
	\includegraphics[width=0.84\linewidth]{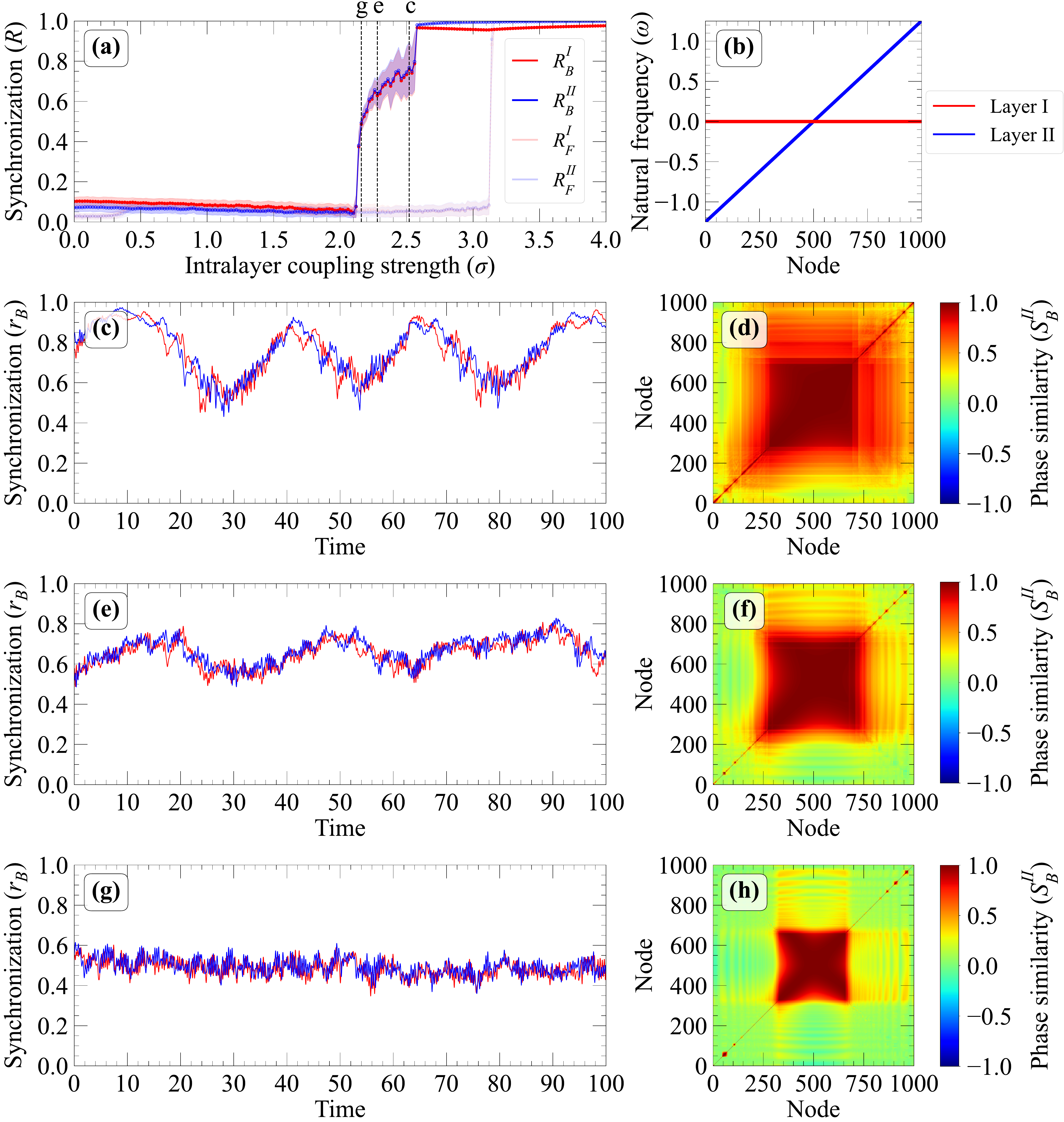}
	\caption{Comprehensive characterization of the backward synchronization transition in a duplex network with two layers, each consisting of $1000$ oscillators. (\textbf{a}) Forward (light curves) and backward (dark curves) synchronization transitions in both layers as functions of the intralayer coupling strength~$\sigma$. Blue and red curves denote the time-averaged synchronization order parameters of layers~I and~II, respectively, while the shaded regions indicate the corresponding temporal standard deviations.
		(\textbf{b}) Natural frequencies of nodes in both layers as functions of the node index for a representative realization with a uniform mirror-node frequency mismatch of width $\delta\omega_{\text{width}} = 2.5$.
		(\textbf{c}, \textbf{e}, \textbf{g}) Time evolution of the synchronization order parameter in both layers for selected coupling strengths $\sigma = 2.52$, $2.28$, and $2.16$, as indicated by dashed lines in panel~(\textbf{a}). The corresponding time-averaged phase similarity matrices of layer~I are shown in the same row in panels~(\textbf{d}, \textbf{f},\textbf{ h}).}
	\label{fig:fig6}
\end{figure}

We first clarify the role of the frustration parameter by considering the interlayer interaction term of the frustrated Kuramoto model, $\sin(\theta_j - \theta_i - \alpha)$. This term can be expressed as a combination of dissipative and reactive components:
\begin{equation}
	\sin(\theta_j - \theta_i - \alpha) = \sin(\theta_j - \theta_i)\cos\alpha - \cos(\theta_j - \theta_i)\sin\alpha.
		\label{eq:11}
\end{equation}
	 For $|\alpha| < \frac{\pi}{2}$, both components are present: the dissipative part reduces phase differences and promotes synchronization, whereas the reactive part primarily induces phase shifts and modulates the collective dynamics. In the limiting case $\alpha = \frac{\pi}{2}$, the dissipative term vanishes ($\cos\alpha = 0$), and the interaction becomes purely reactive, altering the dynamical response of the oscillators without directly enforcing phase alignment.

Previous studies have shown that, in duplex networks, the coexistence of interlayer frequency mismatches and an interlayer frustration parameter close to $\alpha=\frac{\pi}{2}$ can give rise to explosive synchronization in the dynamics of each layer~\cite{pal2025explosive,kachhvah2019delay,kumar2020interlayer}. Moreover, our earlier work demonstrated that not only the average interlayer frequency mismatch but also the specific configuration of frequency mismatches between mirror nodes, plays a crucial role. For a fixed average frequency mismatch ($\Delta\omega$), different configurations of mirror-node mismatches can lead either to explosive synchronization with a stationary order parameter or to double-hysteresis transitions characterized by oscillatory behavior of the order parameter. However, the mechanisms responsible for the emergence of limit cycles and oscillatory behavior of the Kuramoto order parameter in the phase space of duplex networks remained unresolved. Here, we address this gap by systematically identifying the frequency-mismatch configurations that lead to oscillatory behavior, based on a detailed analysis of synchronization dynamics in both layers.

To explore how different frequency configurations influence the phase transition, we assign the natural frequencies of nodes in each layer from Gaussian distributions, constructed so that the frequency differences between corresponding (mirror) nodes follow a uniform, step-function profile. Equivalently, when the nodes are ordered by their frequency mismatches, the magnitude of the mismatch varies linearly with the node index (see Fig.~\ref{fig:fig2}(a, b)). As a second configuration, while preserving the natural frequency distributions within each layer, we randomly reassign the mirror-node pairings. This generates a random mismatch configuration with the same average frequency mismatch as in the first case ($\Delta\omega = 0.56$), but with Gaussian-distributed frequency differences between mirror nodes (see Fig.~\ref{fig:fig2}(d, e)). In the phase transitions of both configurations, we observe explosive synchronization accompanied by a clear hysteresis loop. However, when the frequency-mismatch distribution follows a uniform profile, the backward branch of the phase transition exhibits pronounced fluctuations of the order parameter around its mean value. In contrast, for Gaussian-distributed frequency mismatches, such fluctuations are absent (see Fig.~\ref{fig:fig2}(c, f)). Notably, the only structural and dynamical difference between the two configurations lies in the form of the frequency-mismatch distribution. This observation motivates a detailed examination of the dynamics associated with each configuration to clarify the origin and physical significance of the fluctuations observed along the backward path.

Figure~\ref{fig:fig3} compares the detailed dynamics of the two configurations introduced in the previous figure for a coupling strength within the hysteresis loop ($\sigma = 2.34$). In particular, for the uniform distribution of interlayer frequency mismatches—where fluctuations are observed along the backward transition—the order parameter exhibits sustained temporal oscillations with pronounced amplitudes in the synchronization of both layers. By contrast, such oscillatory behavior is absent for the configuration with Gaussian-distributed interlayer frequency mismatches. In addition, the oscillatory dynamics are composed of multiple coexisting frequencies, indicating that different oscillation modes are superimposed on one another rather than arising from a single dominant frequency. Moreover, at the slowest frequency the oscillations in both layers are in phase, whereas at the second slow frequency the oscillations become anti-phase between the two layers (see Fig.~\ref{fig:fig3}(a)).

To further investigate the local coordinated behavior of the nodes, we plotted the average similarity matrices for
 both layers in the stationary state for each configuration. The node indices in the similarity matrices are ordered according to the magnitude of the frequency mismatches, from low to high. In both configurations, nodes connected to mirror nodes with smaller absolute frequency mismatches show stronger synchronization within their own layers, highlighted by the red squares at the centers of the similarity matrices. However, in the uniform mismatch configuration, a clear gradient from synchronized to incoherent states emerges as one moves from the center toward the peripheral regions of the matrix—a feature that is absent in the Gaussian mismatch configuration. This pattern is observed consistently across both layers, indicating that peripheral nodes—i.e., those with large absolute frequency mismatches with their mirror pairs—are primarily responsible for the observed oscillations.

These oscillations arise because, within each layer, groups of nodes alternate between synchronized and desynchronized states in a “blinking” pattern governed by their frequency mismatches with mirror nodes in the other layer. Specifically, when nodes with large (positive) frequency mismatches become synchronized, those with small (negative) mismatches tend to desynchronize, and this behavior periodically reverses over time, giving rise to the observed oscillatory dynamics. In addition, the size of the blinking groups varies in time, with desynchronization increasing from the peaks to the troughs of the slow-mode oscillations, which explains the synchronization gradient observed from central to peripheral nodes in the similarity matrices. Notably, this dynamical pattern is absent when the interlayer frequency mismatches are Gaussian-distributed. For a more intuitive illustration of these dynamics, see Supplementary Videos S1.

Having established that uniform (step-function) interlayer frequency-mismatch distributions give rise to oscillations composed of multiple modes in the order parameter of each layer, we next examine how the extent of the step function—that is, the range of frequency mismatches between mirror nodes—affects both the nature of the phase transition and the characteristics of the oscillatory dynamics. To systematically control the width of the step function in the frequency-mismatch distribution, we adopt a particularly simple frequency-assignment scheme for the two layers, which allows the mismatch range to be varied in a controlled and systematic manner.
In the duplex network studied here, all natural frequencies in layer~I are uniformly set to zero
($\omega_i^{I} = 0$), while in layer~II the natural frequencies increase linearly with the node index
($\omega_i^{II} = \beta i$). As the slope of this linear profile in layer~II increases, the range of
the uniform distribution of frequency differences between mirror nodes widens, thereby providing
precise control over the width of the interlayer frequency-mismatch distribution. To quantitatively
characterize this width, we define it as the width of the frequency-difference distribution, given by
\begin{equation}
	\delta\omega_{\text{width}}= \max \left( \delta\omega_i\right) -\min \left(\delta\omega_j\right) = \omega_{N}^{II} - \omega_{1}^{II}, \quad i,j = 1,\dots,N,
	\label{eq:12}
\end{equation}

where $N$ denotes the total number of oscillators in each layer. Since the frequencies in layer~II are ordered by the node index, $\omega_1^{II}$ and $\omega_N^{II}$ correspond to the minimum and maximum natural frequencies in that layer, respectively. Hereafter, this measure, $\delta\omega_{\text{width}}$, is referred to as the \textit{interlayer frequency-mismatch width}, and we investigate how variations in this parameter influence both the synchronization transition and the characteristics of the oscillatory dynamics.

Figure~\ref{fig:fig5}(a) shows different configurations of frequency-mismatch assignments corresponding to varying widths of the step function. The red dashed line represents the frequencies in layer~I, which are set to zero for all nodes, while the blue solid lines depict several frequency configurations in layer~II. Figure~\ref{fig:fig2}(b) presents the probability density of frequency differences between mirror nodes, where the color intensity—from light to dark blue—indicates increasing interlayer frequency-mismatch width.

We next systematically examine how the width of the interlayer frequency mismatch influences the phase transitions. Figures~\ref{fig:fig5}(c) and~\ref{fig:fig5}(d) present the forward and backward transition paths, respectively. Since layers~I and~II exhibit qualitatively similar behavior, only the results for layer~II are shown. All simulations were performed on networks of $N = 1000$ oscillators per layer, based on Eqs.~\ref{eq:1} and~\ref{eq:2}, with the interlayer phase frustration fixed at $\alpha = \frac{\pi}{2}$.

Figure~\ref{fig:fig5} highlights the impact of the interlayer frequency-mismatch width on synchronization transitions. In Fig.~\ref{fig:fig5}(c), the forward path displays an abrupt, explosive synchronization, while Fig.~\ref{fig:fig5}(d) shows that the backward transition is also explosive but accompanied by pronounced fluctuations along the path. Importantly, the range of coupling strengths over which these fluctuations occur increases as the width of the interlayer frequency mismatch grows.

As the mismatch width increases, the critical coupling strength along both the forward and backward paths shifts to higher values, reflecting the need for stronger intralayer coupling to achieve synchronization along the forward path and indicating that desynchronization occurs at higher coupling strengths along the backward path (see Fig.~\ref{fig:fig5}(e)). This behavior arises from the increased heterogeneity in natural frequencies caused by the wider interlayer frequency-mismatch distribution, which makes synchronization more difficult~\cite{acebron2005kuramoto}. Despite these changes, the area of the hysteresis loop remains largely unchanged, indicating that the range of coupling strengths over which bistability between synchronized and desynchronized states persists is robust (see Fig.~\ref{fig:fig5}(f)).

We observe that increasing the interlayer frequency-mismatch width broadens the range of coupling strengths over which fluctuations in the order parameter occur along the backward transition path, indicating that oscillations emerge over a wider coupling interval for larger mismatch widths. We now examine in detail how the amplitude and frequency of these oscillations vary with intralayer coupling strength for a fixed interlayer frequency-mismatch width. 

Figure~\ref{fig:fig6}(a) shows the phase transitions and hysteresis loops for both layers, indicated by red and blue lines. The data correspond to the frequency configuration in Fig.~\ref{fig:fig6}(b), with 
$\delta\omega_{\text{width}}=2.5$. Forward transition paths are shown with light lines. Backward paths, where oscillations occur and are of particular interest, are highlighted with dark, bold lines. Shaded areas represent the temporal standard deviations of the order parameter, which increase as the intralayer coupling strength increases within the hysteresis loops.

Figures~\ref{fig:fig6}(c, e, and g) show the time evolution of the order parameter for both layers, indicated by red and blue lines, for three intralayer coupling values marked by dashed vertical lines in Fig.~\ref{fig:fig6}(a). At long timescales (slow modes), the dynamics of both layers are in phase, whereas at shorter timescales they exhibit anti-phase behavior. Furthermore, a comparison across these panels shows that decreasing the intralayer coupling strength leads to a gradual decrease in the amplitude of slow-frequency oscillations within each layer, accompanied by an increase in their wavelength. At small intralayer coupling values within the hysteresis loop, the slow mode disappears.

The time-averaged phase similarity matrices for layer~II are shown in the same row as their corresponding time evolutions for similar intralayer coupling strengths. Since the results for both layers are similar, only the matrices for layer~II are presented. As in Fig.~\ref{fig:fig3}(a), nodes coupled to mirror nodes with similar frequencies tend to synchronize within the layer. At higher intralayer couplings, where pronounced oscillations are observed, the similarity matrices reveal a clear gradient from synchronized to desynchronized groups, progressing from central to peripheral nodes—that is, from nodes with small absolute frequency mismatches with their mirrors to those with larger mismatches. This further confirms that the observed gradient arises from the oscillations. Consistent with the dynamics shown in Supplementary Video~S1, nodes with larger absolute frequency mismatches—typically the peripheral nodes in the similarity matrix—begin to desynchronize at the peak of the oscillation. As the oscillation approaches its minimum, the desynchronized region expands, producing a repeated “blinking” pattern that alternates between synchronized and desynchronized states across the right and left sides of the similarity matrix.

These results reveal a direct connection between the interlayer frequency-mismatch configuration, the emergence of oscillatory dynamics, and the local patterns of synchronization within each layer. The amplitudes of these oscillations and their phase relationships across nodes and layers indicate that the propagation of information within and between layers is strongly influenced by the underlying frequency heterogeneity. To quantify this effect, we now employ an information-theoretic approach, examining directed information transfer both within individual layers and across layers. This analysis provides a detailed characterization of how frequency mismatches shape coordination and information flow throughout the duplex network.


\subsection{Transfer Entropy Analysis}
\label{Transfer Entropy Analysis}
To gain a deeper understanding of the dynamical interactions that govern synchronization, we analyze the network using the transfer entropy (TE) framework, a model-free measure of directed information transfer~\cite{schreiber2000measuring}. TE quantifies how much the past state of a source oscillator reduces the uncertainty of a target oscillator’s future state beyond the information contained in its own history, capturing both the strength and directionality of causal interactions.

In Kuramoto oscillator networks, information transfer between nodes is generally highest near the phase transition points, where the system is partially synchronized and interactions carry the most information, and decreases in both the incoherent and fully synchronized regimes. At these critical points, heterogeneity in natural frequencies and network topology (e.g., node degree, hubs) creates asymmetric influences that enhance directional information flow: higher-frequency nodes tend to lead the dynamics, hubs often act as collective pacemakers sending information to peripheral nodes, and low-degree nodes primarily influence their local neighborhood. Consequently, high-frequency or high-degree nodes typically serve as primary sources of information, whereas other nodes act as sinks. External factors such as time delays, frustration, or interlayer interactions can modify this hierarchy, making previously dominant nodes less predictive and redistributing TE across the network. In the following, we examine how these principles manifest in duplex networks with different types of interlayer coupling.

In the duplex network, interlayer links can be either dissipative ($\alpha = 0$) or reactive ($\alpha = \frac{\pi}{2}$), and their nature significantly shapes the synchronization dynamics and the type of phase transition. Here, we analyze how information flow within each layer reorganizes at the transition points when the interlayer coupling becomes reactive, relating these changes to the observed synchronization behavior. Our results demonstrate that interlayer frustration strongly modulates the pattern of intralayer information transfer, offering a microscopic explanation for the emergence of continuous versus explosive transitions as $\alpha$ varies. We also examine the transfer of information between the two layers in the reactive case ($\alpha = \frac{\pi}{2}$), providing a causal link between macroscopic collective dynamics and the microscopic exchange of predictive information among oscillators.

As mentioned in the Methods section, transfer entropy for the numerical analyses presented here was computed using the Java Information Dynamics Toolkit (JIDT)~\cite{lizier2014jidt}. We employed the multivariate KSG estimator~\cite{kraskov2004estimating}, together with time-shift surrogate data analysis~\cite{theiler1992testing,paluvs2003direction}, to obtain accurate and statistically significant estimates of information transfer. These methods enable quantification of node-level information flow and rigorous significance testing, thereby providing a detailed perspective on the underlying synchronization dynamics.

\begin{figure}[!ht]
	\centering
	\includegraphics[width=0.84\linewidth]{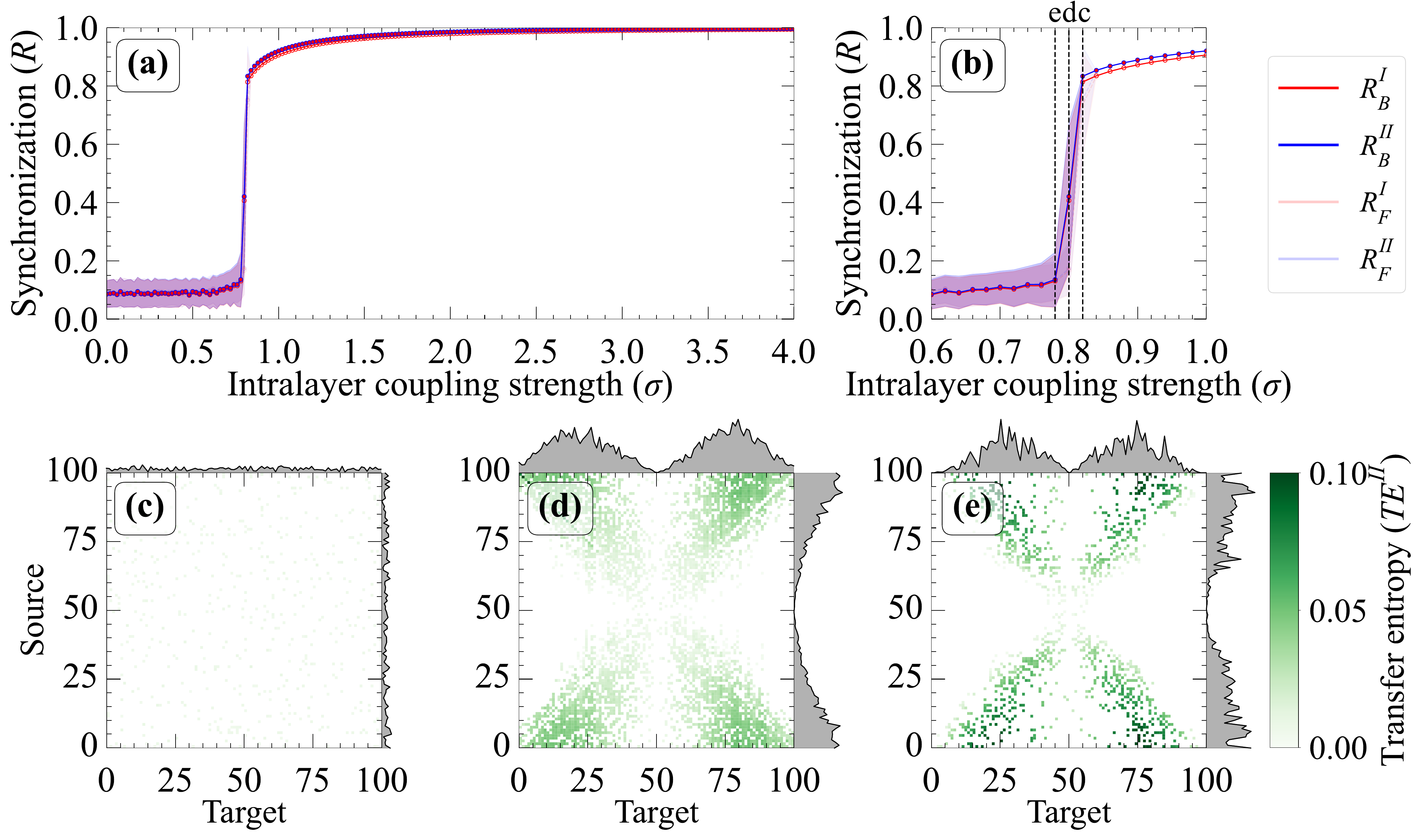}
	\caption{Influence of zero interlayer frustration ($\alpha = 0$) on intralayer synchronization transition and transfer entropy.
		({\bf a}) Synchronization transition in each layer of the duplex network, using a frequency configuration similar to Fig.~\ref{fig:fig5}(a) along the forward and backward paths, with $N = 100$ oscillators per layer and $\delta\omega_{\text{width}} = 2.5$.
		({\bf b}) Enlarged view of panel ({\bf a}). The vertical dashed black lines denote the coupling strengths at which the transfer entropies are computed.
		({\bf c-e}) Heat maps of transfer entropy in layer~II at $\sigma = 0.82$, $0.80$, and $0.78$, respectively, illustrating the backward transition from the synchronized state toward the partially synchronized regime. The marginal curves show the row and column sums, representing the total transfer entropy sent and received by each node.}
	\label{fig:fig7}
\end{figure}

Figure~\ref{fig:fig7} illustrates the synchronization transition and transfer entropy among nodes in the absence of frustration ($\alpha = 0$), where both intralayer and interlayer interactions are purely dissipative. For clarity, we consider the same duplex configuration as in Fig.~\ref{fig:fig6}, where the nodes in layer~I have fixed natural frequencies set to zero, and the natural frequencies of the nodes in layer~II are arranged in ascending order. This arrangement produces a uniform mirror-node frequency mismatch distribution with width $\delta\omega_{\text{width}} = 2.5$. For computational efficiency, we use a duplex network of $N = 100$ oscillators per layer instead of $1000$; this reduction does not affect the dynamics.

Panels~(a) and (b) show that the synchronization transition is continuous, with no hysteresis: the forward and backward branches overlap in both layers, confirming the absence of explosive synchronization when frustration is removed. Panel~(b) highlights the transition region and indicates the coupling strengths used for the transfer entropy analysis in layer~II. Panels~(c–d) present heat maps of the mutual transfer entropy between pairs of nodes, computed at intralayer coupling strengths $\sigma = 0.78$, $0.80$, and $0.82$ along the backward branch of layer~II. These values are close to the critical coupling where the system transitions between synchronized and desynchronized states. The marginal curves alongside the matrices show the row and column sums, representing the total transfer entropy sent from and received by each node. All reported transfer entropy values were statistically significant, passing the time-shift surrogate test with $p < 0.05$.

\begin{figure}[!ht]
	\centering
	\includegraphics[width=1\linewidth]{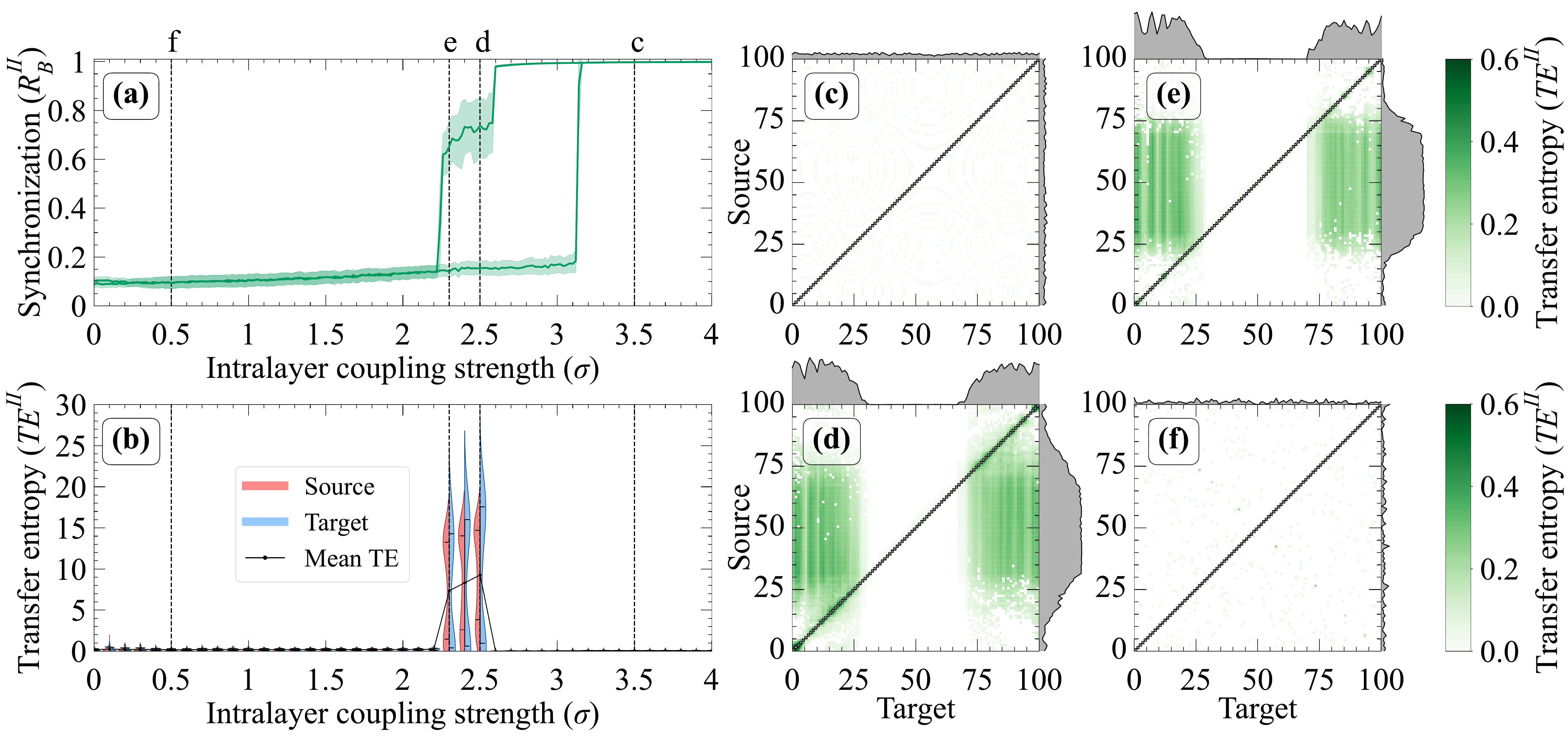}
	\caption{Influence of interlayer frustration ($\alpha = \frac{\pi}{2}$) on intralayer synchronization transition and transfer entropy. The structure and frequency configuration of the network are similar to those in Fig.~\ref{fig:fig7}. (\textbf{a})~Synchronization transition along the forward and backward paths of layer~II as a function of the intralayer coupling strength $\sigma$. Shaded regions represent the temporal standard deviation. (\textbf{b})~Violin plots depicting the distribution of total information sent and received by nodes in layer~II along the backward path, as a function of the intralayer coupling strength $\sigma$. The mean of the distributions is indicated by a solid black curve. (\textbf{c}-\textbf{f})~Pairwise transfer-entropy (TE) matrices for four representative intralayer coupling strengths, corresponding to the dashed vertical lines in panels (\textbf{a}) and (\textbf{b}): $\sigma = 3.50$, $2.50$, $2.30$, and $0.50$, illustrating the transition from coherent to incoherent states. The marginal curves display the row and column sums, representing the total transfer entropy each node sends and receives. }
	\label{fig:fig8}
\end{figure}

As expected, in the coherent state before the transition point, information transfer among node pairs is nearly zero. As coupling increases within the transition region, distinct patterns of information flow emerge in the matrices. Transfer entropy reaches its maximum at the transition points, while the overall flow pattern remains largely consistent throughout the region. Consistent with previous studies, information is primarily transferred according to the ranking of nodes’ absolute frequencies, from higher-frequency nodes to lower-frequency nodes. Moreover, nodes with higher absolute natural frequencies act as stronger information transmitters (sources), whereas the most effective information receivers (targets) are nodes with intermediate absolute natural frequencies, neither very high nor very low.

So far, we have examined the pattern of information transfer between nodes within a layer when the interlayer links are purely dissipative ($\alpha = 0$). We now consider the case where the interlayer interaction is reactive, introducing frustration ($\alpha = \frac{\pi}{2}$), and investigate how this modification alters the information flow within each layer. To this end, while keeping the network structure and frequency configuration the same as in Fig.~\ref{fig:fig7}, we simply add a frustration of $\frac{\pi}{2}$ to the interlayer links. 

Figure~\ref{fig:fig8}(a) shows the synchronization transition along both the forward and backward paths. We observe that reducing the network size to $N = 100$ does not qualitatively affect the dynamics: explosive synchronization with a hysteresis loop and fluctuations along the backward path is still present, similar to the behavior observed for $N = 1000$ in Fig.~\ref{fig:fig6}(a). Thus, the reduced network size, which is important for computational efficiency in calculating transfer entropy, does not alter the essential synchronization dynamics.
\begin{figure}[!ht]
	\centering
	\includegraphics[width=1.0\linewidth]{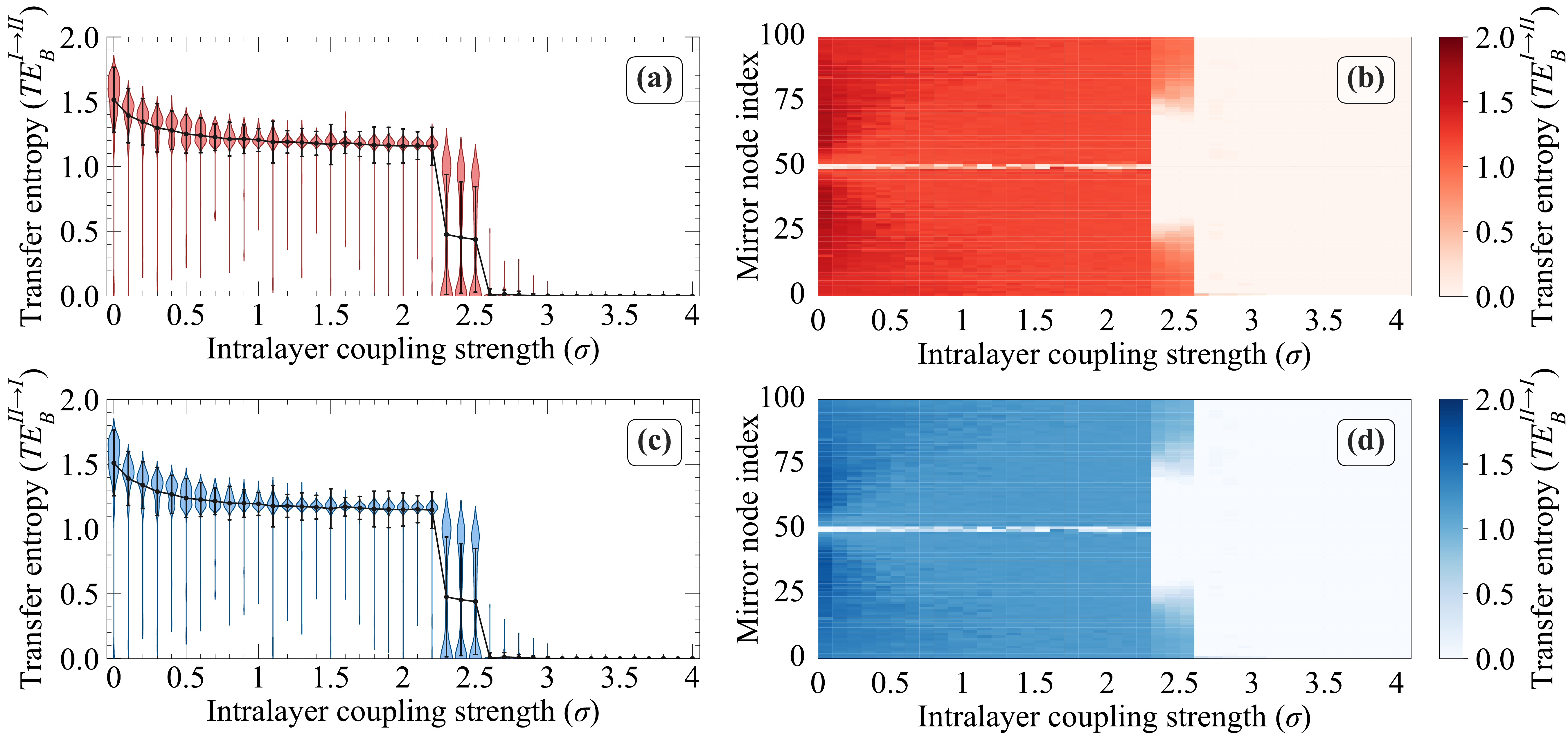}
	\caption{Analysis of information transfer between layers along the backward path in a duplex network with interlayer frustration ($\alpha = \frac{\pi}{2}$). The network structure and frequency configuration are the same as in Fig.~\ref{fig:fig8}. Red and blue indicate information transfer from layer~I to layer~II ($TE^{I \rightarrow II}$) and from layer~II to layer~I ($TE^{II \rightarrow I}$), respectively. (\textbf{a}, \textbf{c}) Violin plots of transfer-entropy distributions across all mirror pairs as a function of $\sigma$, with black dots marking the mean values. Panels (\textbf{b}, \textbf{d)} display color maps of the transfer entropy for individual mirror pairs as a function of intralayer coupling strength. Nodes are ordered according to their interlayer frequency mistmach $\delta \omega_i$, from low to high values.
	}
	
	\label{fig:fig9}
\end{figure}

The violin plots in Fig.~\ref{fig:fig8}(b) show the distribution of total information sent and received by different nodes, computed along the backward path, as a function of intralayer coupling ($\sigma$). As expected, information exchange peaks in the transition region within the hysteresis loop and is nearly zero outside this region. Moreover, the mean transfer entropy increases within the hysteresis region as $\sigma$ grows, indicating that nodes exchange more information at stronger coupling, where the amplitude of the order parameter fluctuations is higher. The distribution of total transfer entropy sent and received by nodes exhibits a bimodal structure, with two distinct peaks corresponding to two groups of nodes that differ in their effectiveness as information sources or receivers.

Figures~\ref{fig:fig8}(c–f) show heatmaps of pairwise transfer entropy between source and target oscillators in layer~II for $\sigma = 3.50$, $2.50$, $2.40$, and $0.50$, respectively, corresponding to the dashed lines in Figs.~\ref{fig:fig8}(a) and (b). As expected, in the weak ($\sigma = 0.50$, panel (c)) and strong ($\sigma = 3.50$, panel (f)) coupling regimes—representing incoherent and fully synchronized states—information transfer between node pairs is negligible, since each oscillator is either independent or fully predictable from its own past, leaving no additional directed information. In contrast, within the transition region ($\sigma = 2.50$ and $2.30$, panels (d) and (e)), the past of one oscillator provides information about the future of another beyond its own history, giving rise to distinct patterns of directed information flow. 

The results further reveal that nodes with smaller frequency mismatches relative to their mirror counterparts—i.e., nodes whose natural frequencies are closer to zero in this configuration—act as stronger information senders within their layers and are thus the most influential. Being already synchronized within their layers, these nodes can drive the dynamics of less synchronized nodes. Therefore, nodes whose frequencies closely match those of their mirrors have the greatest influence, while those with larger mismatches are more strongly affected. For the scenario with $\alpha = \frac{\pi}{2}$, this information flow pattern differs from that observed for $\alpha = 0$ in Fig.~\ref{fig:fig7}, which explains the distinct phase transition behaviors in the two cases. Comparing the two figures, we see that for continuous phase transitions, nodes with larger absolute frequencies serve as the main drivers of the dynamics, whereas for discontinuous transitions, nodes with smaller absolute frequencies are the primary drivers and most influential.

Additionally, the transfer entropy matrices (panels (d) and (e)) show that along the backward path, as the coupling decreases, the number of influential nodes diminishes, eventually leaving only those with the smallest frequency mismatches. This observation is consistent with the results in panel (b), where decreasing the intralayer coupling along the backward path leads to a reduction in the average total transfer entropy. This reduction occurs because the central region of the synchronized cluster—the part that primarily drives the dynamics—shrinks as the intralayer coupling decreases (see heat maps in Fig.~\ref{fig:fig6}). 

Figure~\ref{fig:fig8} illustrates the information flow within layer~II, while Supplementary Fig.~SF1 shows the corresponding flow in layer~I, which is similar to that in Fig.~\ref{fig:fig9}, indicating consistent behavior across layers. Supplementary Fig.~SF2 provides a detailed view demonstrating how the transfer entropy matrices were derived from the surrogate test.

In addition, the distinct patterns in the intralayer transfer entropy matrices—observed for dissipative ($\alpha=0$, Fig.~\ref{fig:fig7}) versus reactive ($\alpha=\frac{\pi}{2}$, Fig.~\ref{fig:fig8}) interlayer links—are primarily determined by the interlayer frustration values rather than the specific frequency arrangement. Similar patterns are observed when reactive interlayer links ($\alpha=\frac{\pi}{2}$) are combined with a Gaussian interlayer frequency mismatch distribution. In this Gaussian case—where fluctuations are absent along the backward path of the synchronization transition—nodes with small frequency mismatches relative to their mirror counterparts not only drive nodes with larger mismatches but also exhibit strong self-information flow within their layers (see Supplementary Fig.~SF3).

Up to now, we have investigated information transfer within layers, both without and with interlayer frustration, i.e., in cases exhibiting continuous or discontinuous phase transitions. Figure~\ref{fig:fig9} summarizes the interlayer transfer-entropy analysis, illustrating how the direction and magnitude of information exchange between mirror nodes vary with the coupling strength.
Figures~\ref{fig:fig9}(a, c) depict the distributions of $TE^{I \rightarrow II}$ and $TE^{II \rightarrow I}$ along the backward path across the full range of intralayer coupling strengths, highlighting the flow of information from layer~I to layer~II and vice versa. The distributions are visualized as violin plots. At high intralayer coupling ($\sigma$), global synchronization suppresses interlayer information transfer and eliminates causal differentiation between the layers. When $\sigma$ decreases to intermediate values—where each layer exhibits a hysteresis loop in its phase transition and nodes exchange information within their own layer (see Fig.~\ref{fig:fig8}(a, b))—we also observe a distribution of interlayer transfer entropies, indicating that mirror nodes across layers exchange information. Notably, the distributions of $TE^{I \rightarrow II}$ and $TE^{II \rightarrow I}$ exhibit a two-peak structure, reflecting distinct patterns of interlayer information transfer, with one group of mirror pairs actively exchanging information and another group not participating in interlayer transfer.

Decreasing $\sigma$ further into the regime of incoherent states—where there is no intralayer information transfer (see Fig.~\ref{fig:fig8}(a, b))—we still observe strong interlayer information exchange between mirror pairs. This occurs because, even at very small $\sigma$ values or $\sigma = 0$, the two layers remain connected through the interlayer coupling strength $\lambda$, allowing mirror nodes to exchange information across layers despite the lack of intralayer synchronization. By increasing $\sigma$ from zero at a fixed $\lambda$, intralayer information exchange is strengthened, and the influence of intralayer coupling becomes evident in the interlayer information exchange between mirror nodes as the system approaches the transition region, as illustrated in Figs.~\ref{fig:fig9}(a, c).

Figures~\ref{fig:fig9}(b, d) show the changes in transfer entropies, $TE^{I \rightarrow II}$ and $TE^{II \rightarrow I}$, for all mirror pairs as a function of $\sigma$. As expected, at high $\sigma$ values—where the layers are fully synchronized—interlayer information flow is negligible for all mirror pairs. Decreasing $\sigma$ to the transition region (hysteresis region in Fig.~\ref{fig:fig8}(a)), mirror pairs with large interlayer frequency mismatch, $\delta\omega_i$, begin to exchange information, while those with small $\delta\omega_i$ still do not, as they remain synchronized within their own layer in this transition regime. These behaviors correspond to the two peaks observed in panels (a) and (c) within the transition regions. Below the transition region, most mirror pairs exchange information, and for smaller $\sigma$ values, the transfer entropies increase, because nodes communicate more easily with their mirrors in the other layer than with their neighbors within the same layer. 

Although the distributions of natural frequencies in each layer were different—one layer set to zero and the other drawn from a uniform distribution—there is no preferred direction for transfer entropy between layer~I and layer~II. This observation suggests that cross-layer interactions are primarily governed by the absolute natural-frequency mismatch of mirror nodes, which serves as the main driving source of influence between the layers.

Finally, a comparison of Figs.~\ref{fig:fig7}, \ref{fig:fig8}, and \ref{fig:fig9} shows that interlayer transfer entropy values are higher than intralayer ones. Moreover, when interlayer frustration is present and interactions are reactive, intralayer transfer entropy values also increase.


\section{Discussion}
\label{Discussion}

In this study, we investigated the dynamics of duplex networks by extending the Kuramoto model to multiplex systems. We focused on how interlayer interactions—ranging from dissipative to reactive (frustrated) links—and the distribution of interlayer frequency mismatches influence collective behavior. In particular, we examined their effects on synchronization transitions, the emergence of oscillatory dynamics, and the patterns of information transfer both within and across layers.

Consistent with previous studies, we found that changing the interlayer links from dissipative ($\alpha = 0$) to reactive ($\alpha = \frac{\pi}{2}$), in combination with interlayer frequency mismatches, can shift the nature of the phase transition from continuous to explosive synchronization~\cite{kumar2021explosive,seif2025double}. In both types of transitions, information flow between nodes peaks in the transition region, with total transfer entropy being higher for explosive synchronization, reflecting stronger and more directed exchanges of information.

For continuous transitions, nodes with the largest absolute natural frequencies act as primary drivers, transmitting information to other nodes, while nodes with intermediate frequencies are the most effective receivers. Nodes near the network’s average frequency contribute minimally to information flow. This hierarchical pattern underlies the gradual and orderly emergence of synchronization.

Introducing reactive interlayer links significantly reshapes intralayer dynamics and the pattern of information transfer, altering which nodes act as drivers or receivers. Each node is simultaneously influenced by intralayer and interlayer couplings. Within the hysteresis region, nodes with small interlayer frequency mismatches synchronize first and act as dominant information sources, driving the dynamics of nodes with larger mismatches. As intralayer coupling $\sigma$ increases, nodes with larger mismatches gradually join the synchronized group, participating in both intra- and interlayer information flow. This process creates conditions that favor the emergence of collective oscillations.

Explosive synchronization thus arises as a macroscopic consequence of the interplay between interlayer frequency arrangements and reactive links, which reorganize intralayer information flow. The emergence of collective oscillations along the backward path of the order parameter is highly sensitive to the distribution of interlayer frequency mismatches. By systematically varying both the profile and width of mirror-node frequency differences, we showed that these factors critically determine the range, amplitude, and modality of oscillatory behavior.

When natural frequencies in the two layers are drawn from different Gaussian distributions, we can construct interlayer arrangements with the same average mismatch but different distributions—Gaussian versus uniform. Both arrangements produce explosive synchronization, but multimodal collective oscillations only appear for uniform interlayer mismatches. While the general flow of intralayer information is similar in both cases—from synchronized cores to more incoherent peripheral nodes—the dynamics differ: Gaussian arrangements yield a large core with peripheral nodes smoothly joining, whereas uniform arrangements involve substantial nodes struggling to synchronize, giving rise to pronounced oscillations.

By controlling the width of uniform interlayer mismatches, we further showed that broader distributions extend the range of intralayer coupling $\sigma$ over which the order parameter oscillates; additionally, increasing $\sigma$ enhances the amplitude of these oscillations. These oscillations are multimodal: the slowest mode involves in-phase oscillation of order parameters across layers, while the second slow mode shows antiphase relations. The superposition of these modes—combining central-peripheral synchronization with local diffusion among peripheral nodes—produces rich, blinking dynamics, which may provide insight into how ubiquitous multimodal oscillations, such as those observed in brain networks, are organized.

In conclusion, our results demonstrate that interlayer link frustration and the distribution of interlayer frequency mismatches jointly govern phase transitions and the emergence of oscillatory dynamics. By connecting macroscopic synchronization and multimodal oscillations to microscopic directed information flow, this work provides insight into how complex patterns can emerge from network interactions in multiplex systems.


\section*{Data availability}
\phantomsection
\label{Data availability}
\addcontentsline{toc}{section}{Data availability}
The datasets generated and analysed during the current study are available in the Github repository, \href{https://github.com/Articles-data/seif2026Synchronization}{https://github.com/Articles-data/seif2026Synchronization}.

	\section*{Author contributions}
	\phantomsection
	\label{Author contributions}
	\addcontentsline{toc}{section}{Author contributions}
	M.Z. designed the study and developed the initial concept as a supervisor. A.S. conducted the simulations, performed the analysis, and prepared the results. Both authors discussed the results and contributed to writing the manuscript.

	
	\section*{Additional information}
	\phantomsection
	\label{Additional information}
	\addcontentsline{toc}{section}{Additional information}
	
	{\bf Correspondence} and requests for materials should be addressed to M.Z.

\FloatBarrier

\newpage

\makeatletter
\renewcommand\@biblabel[1]{\textbf{#1.}\hfill}
\makeatother

\maketitle 
\pagenumbering{arabic} 

\setcounter{page}{1} 
\renewcommand{\thepage}{S\arabic{page}} 

\newpage

\section*{
	\begin{center}
		{\fontsize{12}{12}\selectfont Supplementary materials for:}\\
		Synchronization, Collective Oscillations, and Information Flow in Duplex Networks
	\end{center}
}
	
	This supplementary material provides explanations for supplementary figures \textcolor{blue}{SF1-SF3} and supplementary videos \textcolor{blue}{SV1}.

	\section*{Description of Supplementary Figures }
	\phantomsection
	\label{Description of Supplementary Figures}
	\addcontentsline{toc}{section}{Description of Supplementary Figures}
	\subsection*{Description of Figure SF1:}
	\phantomsection
	\label{Figure S1}
	\addcontentsline{toc}{subsection}{Figure S1}
	
	Supplementary Fig.~SF1 compares the transfer entropy matrices for the two layers of the duplex in the case of reactive interlayer couplings ($\alpha = \frac{\pi}{2}$, Fig.~7). The first row shows the intralayer transfer entropies for layer~I, while the second row displays the corresponding matrices for layer~II at the same coupling strengths $\sigma$. The selected coupling strengths for the transfer entropy calculations—identical to those in Fig.~7—are, from left to right, $\sigma = 3.5$, $2.5$, $2.3$, and $0.5$, representing the transition from coherent to incoherent states along the backward path of the phase transition. As expected, there is no information flow between nodes within each layer in the coherent and incoherent states, whereas a clear pattern of information transfer emerges in the transition region corresponding to the hysteresis loop. The pattern of information flow within both layers is similar, originating from synchronized groups of nodes with small frequency mismatches with their mirrors and directed toward nodes with larger mismatches relative to their corresponding mirrors.
	
	\begin{figure}[!ht]
		\renewcommand{\figurename}{Figure SF}
		\centering
		\setcounter{figure}{0}
		\includegraphics[width=1\linewidth]{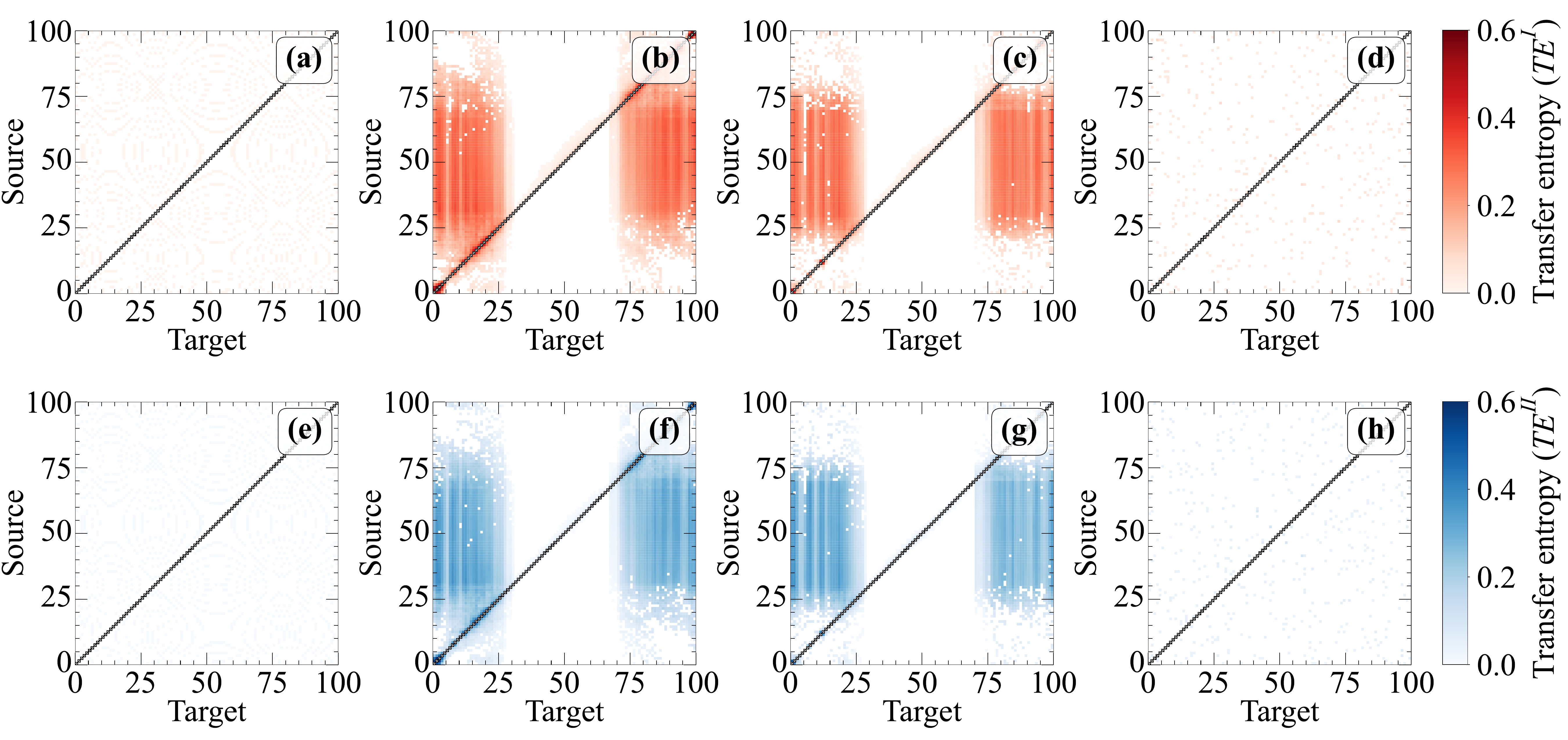}
		\caption{Comparison of intralayer transfer entropy matrices for layer~I (top row, red) and layer~II (bottom row, blue) along the backward path of the phase transition at different coupling strengths. From left to right, $\sigma = 3.5$, $2.5$, $2.3$, and $0.5$, corresponding to the progression from coherent to incoherent states.}

		\label{fig:Fig-S1}
	\end{figure}

\newpage
	\subsection*{Description of Figure SF2:}
	\phantomsection
	\label{Figure S4}
	\addcontentsline{toc}{subsection}{Figure SF4--SF7}
	
	Supplementary Fig.~SF2 illustrates the procedure used to extract the transfer entropy matrices. Transfer entropy was computed between selected nodes, and a surrogate data test was applied to assess statistical significance. Only values passing the surrogate test ($p < 0.05$) are included in the matrices, ensuring that the measured information flow reflects real interactions rather than random or accidental correlations. 
	
	The figure illustrates the procedure used to extract the transfer entropy matrices reported in Fig.~7 of the main text for $\sigma = 2.30$ (first row) and $\sigma = 0.5$ (second row). Each row consists of three panels arranged from left to right. The left panel shows a heat map of the transfer entropy matrix, where the color intensity (red scale) represents the magnitude of information transfer between pairs of nodes in layer~II of the network. The middle panel displays a heat map obtained from a surrogate data analysis. Statistical significance was assessed by generating 500 surrogate realizations through random circular time shifts; transfer entropy values with $p < 0.05$ are considered significant, while non-significant values are discarded in the filtered analysis. The right panel (green scale) shows the filtered transfer entropy matrix used in the paper, where only statistically significant values are retained and all other entries are set to zero. In all panels, the vertical axis corresponds to source nodes (outgoing information), and the horizontal axis corresponds to target nodes (incoming information).

	\begin{figure}[!ht]
		\renewcommand{\figurename}{Figure SF}
		\centering
		\includegraphics[width=1\linewidth]{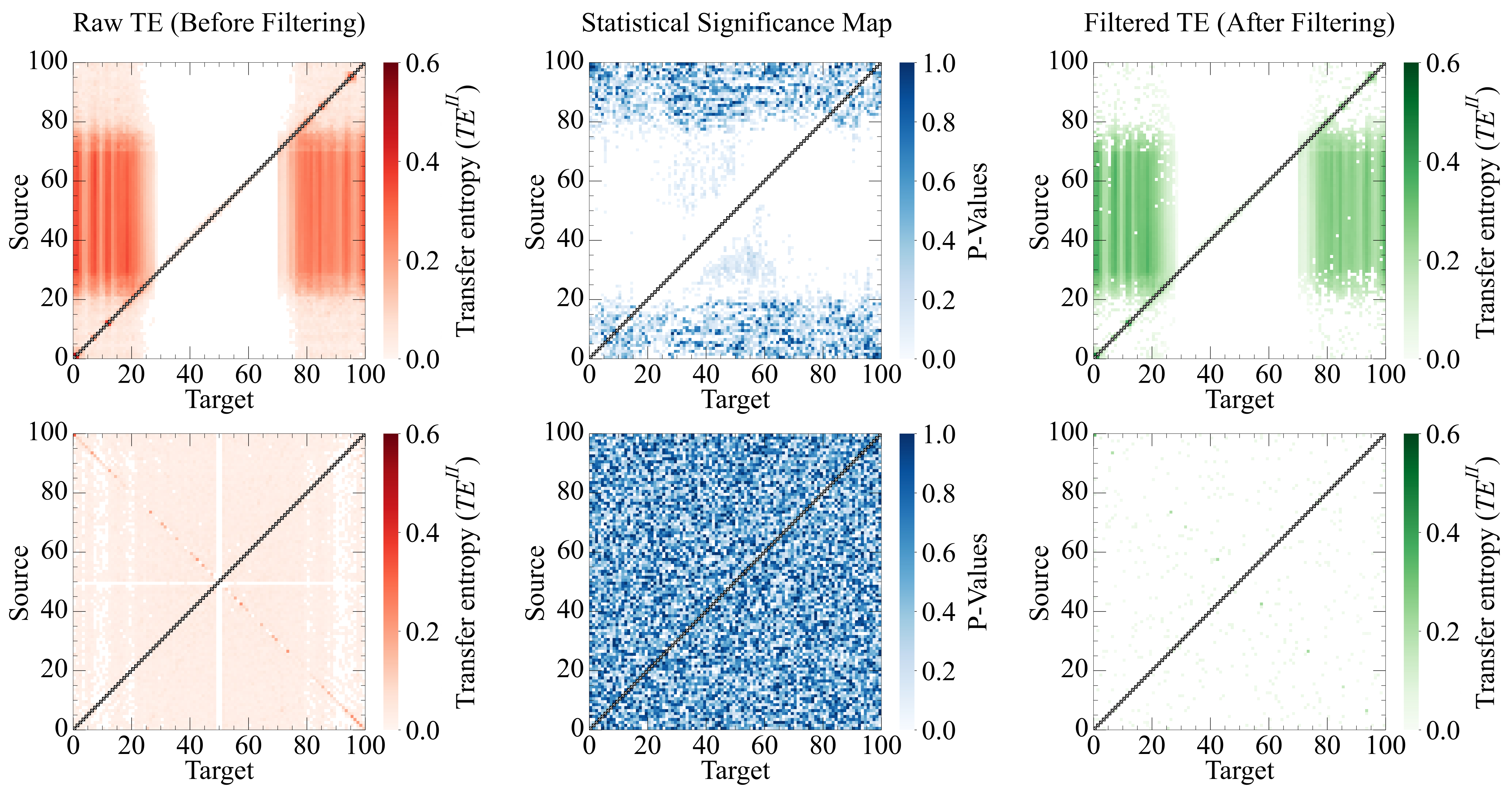}
		\caption{ Procedure used to extract the transfer entropy matrices reported in Fig.~7 of the main text for $\sigma = 2.30$ (first row) and $\sigma = 0.5$ (second row). Matrices in each row are shown from left to right: raw transfer entropy, surrogate analysis, and surrogate-filtered transfer entropy.}
		
		\label{fig:Fig-S4}
	\end{figure}

	\newpage
	\subsection*{Description of Figure SF3:}
	\phantomsection
	\label{Figure S3}
	\addcontentsline{toc}{subsection}{Figure S3}
	Supplementary Fig.~SF3 compares the intralayer dynamics and information transfer matrices of duplex networks under two distinct interlayer frequency mismatch distributions: a uniform (first row) distribution and a Gaussian (second row) distribution. In both cases, the interlayer frustration is set to $\alpha = \frac{\pi}{2}$, and the network structure and frequency arrangements are identical to those in Fig.~2 of the main text.
	
	Due to computational limitations in estimating transfer entropies for all pairs of nodes in layer~II, we first sort the nodes according to the intrinsic frequency differences with their corresponding mirror nodes, $\delta\omega_i$, in ascending order. We then select 100 nodes at equal intervals from the 1000 nodes in layer~II and compute the transfer entropy matrices among them.
	
	For each case (uniform and Gaussian distributions), the transfer matrices are shown for two intralayer coupling strengths within the hysteresis loop, as indicated by the dashed lines in panels (a) and (d).

	In the uniform (step-function) case, similar to the scenario explained in Fig.~7 of the main text (obtained using a different algorithm but yielding the same uniform frequency mismatch distribution), the information transferred from the synchronized groups of nodes—those with small frequency mismatches relative to their mirror nodes—to the nonsynchronized groups of nodes—those with large frequency mismatches relative to their mirror nodes—is predominantly directed from the former to the latter. There is also information flow within the group of unsynchronized nodes with large absolute frequency differences relative to their mirror nodes, particularly for large $\sigma$ values within the hysteresis loop, where the amplitude of the order parameter oscillations is higher. This intragroup information decreases as $\sigma$ decreases. Consequently, for a uniform interlayer frequency mismatch distribution, the total information flow—the sum of the transfer entropy matrices—decreases along the hysteresis loop as $\sigma$ decreases: $TE^U_{\text{total}} = 720.61$ at $\sigma = 2.34$, and $TE^U_{\text{total}} = 555.77$ at $\sigma = 2.06$.

	For the case with a Gaussian frequency mismatch distribution, in contrast to the uniform case, the total information flow between nodes within a layer increases as $\sigma$ decreases within the hysteresis loop region ($TE^G_{\text{total}} = 322.56$ at $\sigma = 2.34$, and $TE^G_{\text{total}} = 573.17$ at $\sigma = 2.06$). There is only a small flow of information between nodes at $\sigma = 2.34$, where large groups of nodes are well synchronized (see Fig.~3). At $\sigma = 2.06$, similar patterns of information flow as in the uniform distribution case appear, with information predominantly directed from nodes with small absolute frequency mismatches to those with large absolute frequency mismatches. However, the information flow among nodes within the group with large interlayer frequency mismatches is very limited, which prevents the generation of large-amplitude oscillations in the order parameter.
	
	\begin{figure}[!ht]
		\renewcommand{\figurename}{Figure SF}
		\centering
		\includegraphics[width=1\linewidth]{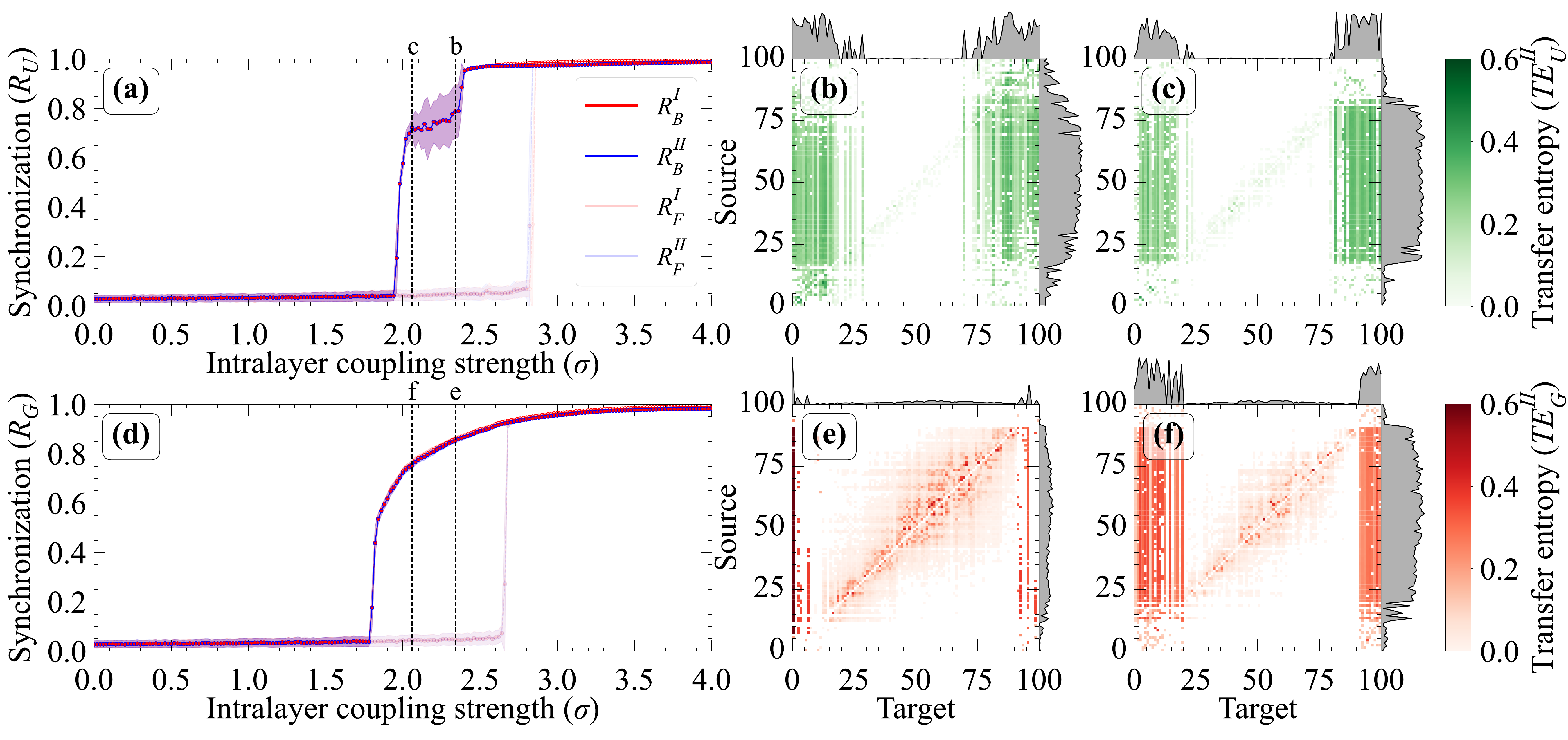}
		\caption{Comparison of the dynamics and transfer entropy matrices for layer~II of a duplex network with $\alpha = \frac{\pi}{2}$ under two different interlayer frequency mismatch distributions: uniform (first row) and Gaussian (second row).
			Panels ({\bf a, d}) and show the synchronization transition as a function of intralayer coupling strength for each case. The heat maps shown in panels ({\bf b, c}) and ({\bf e, f}) illustrate, for each case, the transfer entropy matrices in layer~II along the backward paths at $\sigma = 2.34$ and $\sigma = 2.06$, as indicated by the dashed lines in panels ({\bf a, d}). The network structures and frequency arrangements are identical to those in Fig.~2 of the main text. However, the transfer entropy matrices are calculated for $100$ nodes selected at equal intervals from the $1000$ nodes in layer~II. The nodes are ordered according to the intrinsic frequency differences with their corresponding mirror nodes, $\delta\omega_i$, in ascending order. }
		
		\label{fig:Fig-S3}
	\end{figure}

	\newpage
	\section*{Description of Supplementary Videos}
	\phantomsection
	\label{Description of Supplementary Videos}
	\addcontentsline{toc}{section}{Description of Supplementary Videos}
	Selected snapshots from the supplementary video are presented in this section to aid the interpretation of the panels.
	
	\subsection*{Description of video SV1}
	\phantomsection
	\label{Figure SV1}
	\addcontentsline{toc}{subsection}{Figure SV1}
	Snapshot from a video showing how the interlayer frequency mismatch distribution shapes the dynamical behavior within each layer, enabling a comparison between the Gaussian and uniform profiles. Both distributions exhibit explosive synchronization due to the reactive interlayer interactions ($\alpha = \frac{\pi}{2}$). However, clear oscillations in the order parameter are observed in the uniform distribution, whereas they are absent in the Gaussian distribution. The video illustrates in detail how these oscillations arise from the coordinated “dancing” of the phases of oscillators with large interlayer frequency mismatches relative to their mirrors on the unit circle (peripheral nodes in the similarity matrices in the second row), a phenomenon referred to as the blinking process in the main text.
	\begin{figure}[!ht]
		\renewcommand{\figurename}{Figure SV}
		\centering
			\setcounter{figure}{0}
		\includegraphics[width=1\linewidth]{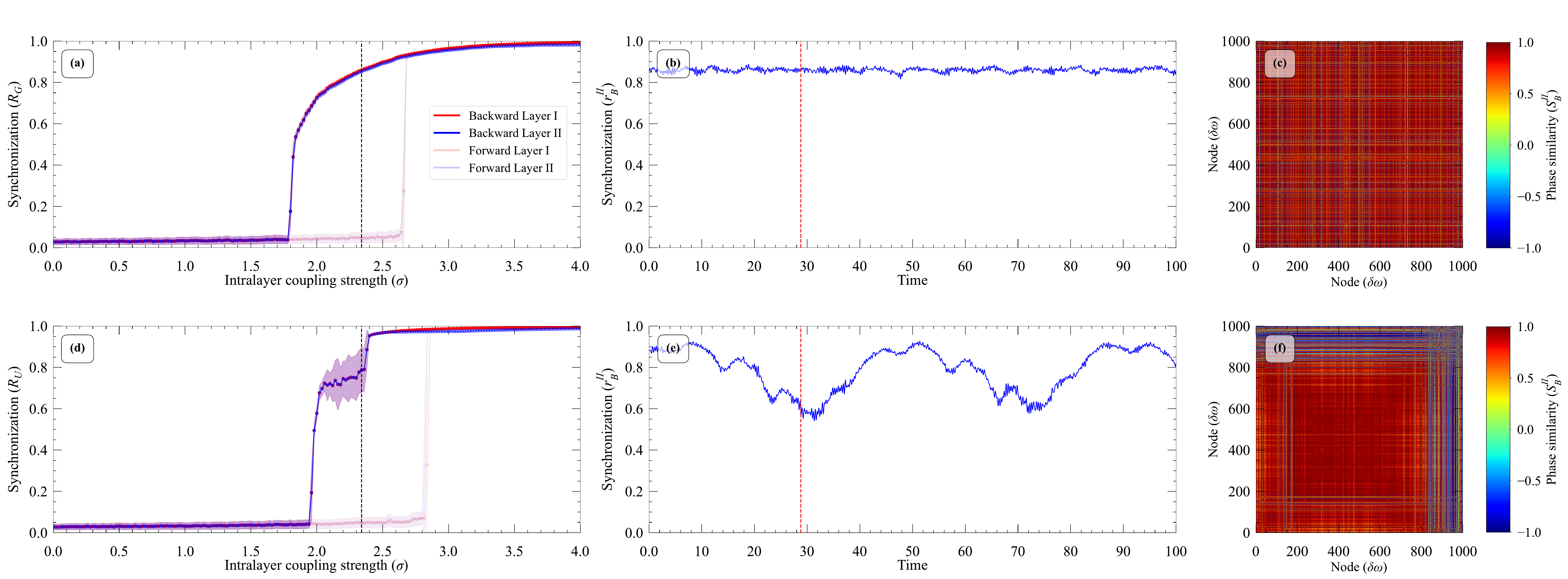}
		\caption{Snapshot from a video comparing the dynamics of layer~II in a duplex network with $\alpha = \frac{\pi}{2}$ under two different interlayer frequency mismatch distributions: Gaussian (first row) and uniform (second row). Panels ({\bf a}) and ({\bf d}) show the synchronization phase transition as a function of the intralayer coupling constant $\sigma$ for the forward (light curves) and backward (dark curves) paths. Red curves correspond to layer~I, and blue curves to layer~II. The vertical black dashed line indicates the coupling value at which the dynamical evolution is examined in the video ($\sigma = 2.34$). Panels ({\bf b}) and ({\bf e}) show the temporal evolution of the synchronization order parameter of layer~II at $\sigma = 2.34$, along the backward path. The vertical red dashed lines indicate the time instants at which the similarity matrices are extracted and displayed in panels ({\bf c}) and ({\bf f}). The network structures and frequency arrangements are identical to those in Fig.~2 of the main text, and the nodes are ordered according to the intrinsic frequency differences with their corresponding mirror nodes, $\delta\omega_i$, in ascending order. }
		\label{fig:Fig-SV1}
	\end{figure}

\end{document}